\documentclass[10pt, conference, compsocconf]{IEEEtran}
\usepackage{amsmath}
\usepackage{amsfonts}
\usepackage{setspace}
\usepackage{array}
\usepackage{amssymb}
\usepackage{graphicx}
\usepackage{url}
\usepackage{multicol}
\usepackage{stfloats}
\usepackage{subfigure}
\usepackage[usenames]{color}
\usepackage{flushend}
\usepackage{booktabs}
\usepackage{changes}
\usepackage{eurosym}
\usepackage{rotating}

\newcommand{\bi}{\begin{itemize}}
\newcommand{\ei}{\end{itemize}}
\newcommand {\beq}{\begin{equation}}
\newcommand {\eeq}{\end{equation}}
\newcommand {\be}{\begin{enumerate}}
\newcommand {\ee}{\end{enumerate}}

\begin{document}

\title{Characterization of Cross-posting Activity for Professional Users across Facebook, Twitter and Google+}

\author{\IEEEauthorblockN{Reza Farahbakhsh\IEEEauthorrefmark{1},
\'Angel Cuevas\IEEEauthorrefmark{1}\IEEEauthorrefmark{2} and
No\"{e}l Crespi\IEEEauthorrefmark{1}}
\IEEEauthorblockA{\IEEEauthorrefmark{1}Institut Mines-T\'el\'ecom, T\'el\'ecom SudParis,CNRS UMR 5157 SAMOVAR, France\\
\{reza.farahbakhsh, noel.crespi\}@it-sudparis.eu}
\IEEEauthorrefmark{2}Universidad Carlos III de Madrid,
{acrumin@it.uc3m.es}\\


}

\maketitle
\begin{abstract}

Professional players in social media (e.g., big companies, politician, athletes, celebrities, etc) are intensively using Online Social Networks (OSNs) in order to interact with a huge amount of regular OSN users with different purposes (marketing campaigns, customer feedback,  public reputation improvement, etc). Hence, due to the large catalog of existing OSNs, professional players usually count with OSN accounts in different systems. In this context an interesting question is whether professional users publish the same information across their OSN accounts, or actually they use different OSNs in a different manner. We define as cross-posting activity the action of publishing the same information in two or more OSNs. This paper aims at characterizing the cross-posting activity of professional users across three major OSNs, Facebook, Twitter and Google+. To this end, we perform a large-scale measurement-based analysis across more than 2M posts collected from 616 professional users with active accounts in the three referred OSNs.
Then we characterize the phenomenon of cross posting and analyze the behavioral patterns based on the identified characteristics.\footnote{This is an extended version of a published paper at Asonam'15 conference \cite{asonam_15}}
\end{abstract}



\section{Introduction}
\label{sec:intro}

Online Social Networks (OSNs) have become one of the most popular services in the Internet attracting billions of subscribers and millions of daily active users. This tremendous success has created a very profitable market in which major OSN players have acquired an important role on the current Internet. We can find three dominant OSNs according to their number of subscribers: Facebook (FB), Twitter (TW) and Google+ (G+). While these systems have been demonstrated to be very attractive to regular users that perform a wide variety of social interactions on them, they also present a golden opportunity to professional players\footnote{Through this paper, the term ``professional users'' stand for users in social media that behind their accounts there are entities with a clear business plan that are utilizing social media for their business interest.} (i.e. brands, politicians, celebrities, etc.) to interact with a huge amount of potential customers/voters/fans to increase their reputation and popularity, to run marketing campaigns, to attract voters, etc. In a nutshell, we can find a number of professional players or users that are using OSNs with a similarly professional goal.

Most professional users do not limit their activity to a single OSN, but usually they have accounts in multiple OSNs, including the most popular ones such as FB, TW and G+. Then an interesting question is whether professional players use all OSNs in the same way, or actually they use each OSN for different purposes. In other words, when a professional user wants to advertise or notify some update, does she publish that information in several OSNs?, or contrary, she publishes it in a single OSN depending on the type of information (e.g., if it is a personal update she publishes a post in one OSN, but in case it is a commercial update she selects another OSN). We refer to the information that a professional player publishes in multiple OSNs as \textit{cross-posting activity}. Therefore, if a professional user publishes a post in FB and a post TW that contain the same information we consider them as a cross-post.

To the best of our knowledge, although there are other works that have analyzed the behaviour of regular users across two OSNs \cite{Of_Pins_Tweets,Archival}, this paper presents the first large scale study on cross-posting activity of professional users across the three major OSNs, i.e., FB, TW and G+. We analyze the activity of 616 (popular) professional users with active accounts in the three referred OSNs. Among these users we can find big companies, politicians, athletes, artists, celebrities, public institutions, etc. To perform the study we have analyzed more than 2M posts distributed across the 616 users in TW, FB and G+.

The first contribution of this paper is a simple yet efficient methodology that is able to precisely determine whether two posts contain the same information, and thus classify them as a cross-post. This methodology relies in a hierarchical algorithm implemented in two steps. The first step applies NLTK \cite{ntlk_fuzzy} to compare a pair of posts in a fuzzy mode, and provides a binary decision on whether they actually represent a cross-post. Those pairs of posts obtaining a positive comparison are already classified as cross-post at this stage, while the pairs failing in this comparison go to the second step of the algorithm. The second step of the algorithm uses two metrics, cosine similarity \cite{cosine_similarity} and string similarity \cite{fast_string}, which provides a similarity value ranging between 0 and 1 for each pair of posts under comparison. Then the closer the similarity value is to 1 the more similar the posts are. We classify as cross-post any pair of posts obtaining a similarity value $\geq 0.5$ for both metrics, cosine similarity and string similarity. The validation of our methodology shows an accuracy of 99\% for the classification of cross-posts.

Based on this methodology, the first goal of the paper is to characterize the cross-posting activity of professional OSN users across FB, TW and G+. In order to achieve this objective we perform a  data analysis that allows us to shed light to three key aspects of the cross-posting activity. $(i)$ The first immediate question is whether the cross-posting phenomenon actually exists, and if it exists what fraction of the activity from a professional user is associated to cross-posting.  $(ii)$ In case the cross-posting activity is relevant,  we aim at understanding between which OSNs it is more frequent. This means, can we find more cross-posts between FB-TW, FB-G+, or TW-G+? $(iii)$ Finally, we measure what is the benefit, if any, (in terms of attracted engagements) that professional users obtain from the cross posting activity in compare to non cross posts.

Once we have characterized the cross-posting behaviour, we study which is the preferred OSN of professional users as initial source to inject information. Indeed, when a professional user decides to publish her updates first in an OSN than other, she is privileging the first OSN that somehow is showing the ``breaking news" for that user.

Finally, our last effort defines cross-posting behavioural patterns for users with some representative characteristic that determines their profile.  First, we characterize the behaviour of professional users with a strong preference for initiating their cross-posts in a particular OSN using three metrics: $(i)$ similarity of their cross-posts, $(ii)$ type of content associated to the cross-post they publish, and $(iii)$ sites more frequently linked by the urls contained in their cross-posts. In addition, we repeat the analysis but classifying the users based on their median inter-posting interval, which refers to the time gap between the moment they publish the cross-post in the first OSN and the instant it is uploaded in the second OSN.

In the following, we list the main findings of our research:

\noindent{(1) Cross-posting is a frequent practice across professional users. In median a professional user share in other OSN 25\% of the posts published in FB and G+, and only 3\% of the tweets. However, we must note that professional users are much more active in TW than FB and G+, hence, in absolute terms, the TW account of professional users generate a larger volume of cross-posts than G+ accounts and similar volume to FB accounts.}

\noindent{(2) The cross-posting phenomenon mainly happens between FB and TW, but it is also relevant between FB and G+. However, it is surprising that is more likely to find a cross-post published in FB, TW and G+, than only in TW and G+. Therefore, professional users do not find any benefit on sharing information between their TW and G+ accounts.}


\noindent{(3) Professional users attract a substantial engagement in FB and TW when they publish cross posts since they attract 30\% and 100\% more engagement as compared to non-cross-post. However, in the case of G+ non-cross-posts attract 2$\times$ more engagement than cross-posts.}

\noindent{(4) Among the 616 analyzed users 50\% prefer FB as most frequent option to initially upload their cross-posts, 45\% prefer TW, and only 5\% give priority to G+.}

\noindent {(5) Professional users with a strong preference for TW publish cross-posts that: $(i)$ are very similar across the different OSNs, $(ii)$ mostly includes textual content, and $(iii)$ mostly include links to websites different than OSNs sites.

\noindent {(6) Professional users with a strong preference for FB publish cross-posts that: $(i)$ mostly includes audiovisual content, and $(ii)$ mostly include links to content stored in major OSNs sites.}

\noindent {(7) As the inter-posting interval decreases: (i) the similarity of cross-posts increases, (ii) the portion of audiovisual content attached to cross-posts decreases, (iii) and a larger portion of urls included in cross-posts refers to major OSNs sites.}


\section{Data Collection Methodology}
\label{sec:methodology}

This section briefly explains our data collection methodology to construct the required dataset to achieve the objectives addressed in this paper.


Our first challenge was to identify a numerous group of relevant professional users having active and popular accounts across FB, TW and G+. To this end, we rely on a large dataset that includes thousands of professional and regular users with an account in the three OSNs collected for a previous work \cite{H2H_report}. 
From these users we were interested in those ones that meet two requirements: (i) have an active account in FB, TW and G+; (ii) present a high popularity in at least two of the systems. We found 616 professional users that satisfy the popularity requirement.
We validated that the selected users were actually relevant in all the three OSNs by means of an external source \cite{Socialbakers} ranks professional users in each system in terms of popularity. In addition we validated that each of the selected account belong to same users in three systems. To this end, we used the feature of user verification which allows the owner of the account verifies the validity of the page and this feature is exist in all the three systems. We implemented a script that check each of the accounts in three systems and ensure that it is verified by the owners.
Following we briefly introduce the crawlers developed to retrieve the activity of the professional users from each OSN. For more details on these crawlers we refer the reader to \cite{H2H_report,gonzalez2012google+}:

\textbf{\textit{FB crawler.}}
The crawler receives a user ID (or username) as input and uses the FB API to collect the posts published by the user in her FB account. The API provides quite a lot information from a post from which the most relevant for our paper is: $(i)$ the description of the post that refers to the text included by the user in that post, $(ii)$ the timestamp associated to the exact publication time of the post, and $(iii)$ the type of content associated to the post, which could be photo, video, link (when the post includes an url) and status (that refers to the post that only include text). It must be noted that FB API imposes a maximum threshold of 600 queries every 10 minutes. Hence, in order to speed up our data collection process, we used multiple instances of the crawler working in parallel.

\textbf{\textit{TW crawler.}}
The crawler receives as input a user identifier that can be either the user's id or the user's screen name and queries the Twitter API to obtain the user's profile attributes, the total number of published tweets, and the last 3,200 tweets posted by the user along with the number of reactions associated with each one of the user's tweets, except the responses (i.e., comments) for a tweet. Consequentially if a user has published more than 3,200 tweets we can only retrieve the last 3,200. Twitter imposes a limit of 150 requests per hour per IP address. To overcome this limitation, we use  PlanetLab~\cite{chun2003planetlab} infrastructure to parallelize our data collection process. Specifically, our crawler sends requests to TW API using approximately 450 PlanetLab machines as proxies, so that we can multiply the speed of our data collection in proportion to the number of used proxies.

\textbf{\textit{G+ crawler.}}
This crawler is composed by two modules. The first one collects the public profile information as well as the connectivity information of all the users in the largest connected component (LCC) of G+. This module is a web-crawler that parses the web page of G+ users to collect the previous information. The second module uses the G+ API  to collect all the public posts as well as their associated reactions. Google limits the number of queries to the G+ API to 10K per hour per access token. In order to overcome this limitation we have created several hundred accounts with their correspondent access tokens and leverage the proxies infrastructure in PlanetLab explained above to speed up our crawling data collection.\\

\begin{table}[t]
  \vspace{-1mm}
  \centering
  \scriptsize
  \vspace{-1mm}
  \caption{Datasets description}
  \vspace{-0.25cm}
    \begin{tabular}{cccccc}
    \hline
       OSN & \#users & total \#posts & avg. \#posts per user \\
    \hline
    FB    & 616 & 422 K  & 685 \\
    G+    & 616 & 173 K  & 280 \\
    TW   & 616 & 1.64 M  & 2664 \\
    \hline
    \end{tabular}%
  \label{tab:dataset_info}%
  \vspace{-3mm}
\end{table}%


Table \ref{tab:dataset_info} summarizes the datasets used in this paper. In total, we analyze more than 2M posts published across 616 professional publishers in FB, TW and G+. 
Finally, it must be noted that the collection campaign finished on May 2013, thus our dataset may not include novel features released by any of the analyzed OSNs after that period.

\section{Methodology to Identify Cross-posts}
\label{sec:Cross_posts_methodology}

In order to being able to compare cross-posting activity of professional users we need to have an accurate mechanism that detects when two posts are actually containing the same information based on their posts' textual descriptions but intelligent enough to identify similar posts with additional information in one of the systems.
For instance, a given user could upload a post in FB and TW which refers exactly to a recent event, but in the case of FB she uploads a picture and in TW she adds a link to the picture. In other words, a tool that only detects as cross-posts those posts that are exactly the same in two OSNs is inaccurate for our research. Hence, we have implemented a hierarchical classification algorithm that determines whether two posts can be considered as cross-posts in two steps. Then, given the description (i.e. the text associated to a post) of two posts, $P_1$ retrieved from the account of user $U$ in $OSN_A$ and $P_2$ published by $U$ in her account of $OSN_B$, our algorithm proceeds as follows:

\textbf{(1)} We compare $P_1$ and $P_2$ using NLTK toolkit \cite{ntlk_fuzzy} that provides a binary decision based on the similarity of the compared texts. NLTK Fuzzy Match generates a positive answer (i.e., same text) when both texts are fully similar. Therefore, in the context of cross-posting analysis if NLTK Fuzzy Match determines that $P1$ and $P2$ are similar, we can safely classify them as cross-post. However, in the case that the output is negative we cannot guarantee that $P_1$ and $P_2$ are not referring to the same information, thus we cannot classify them as non-cross-post. In summary, all the pairs of posts receiving a positive classification are labeled as cross-posts while the remaining pairs need to go through the second step of our algorithm.

\textbf{(2)} We compare $P_1$ and $P_2$ using two other similarity methods: cosine similarity technique \cite{cosine_similarity} and string similarity method \cite{fast_string}.
In summary, string matching compares two strings word by word and the result is a value that reveals the percentage of similarity between both strings according to the number of words appearing in both texts. More number of common words will lead to a higher percentage of similarity. This method is independent of the language and can compare texts and words in any language.
On the other hand, cosine similarity method is useful to compare strings that include urls. This complementary similarity check helps us to identify similar texts that just have some additional info on them.
These two methods provide as output a value ranging between 0 and 1, so that the closer is the output to 1 the more similar $P_1$ and $P_2$ are. Based on the obtained results, we classify $P_1$ and $P_2$ as cross-post if both metrics, cosine similarity and string similarity, are $\geq$ 0.5. Later in this section we validate our methodology and demonstrate why we have selected the 0.5 threshold.

The previous algorithm serves to classify any pair of posts as cross or non-cross based on their description. In addition, we must note that our algorithm is not bound to any particular alphabet, so it can be applied in multiple languages. However, the use of the hierarchical algorithm is not enough for the purpose of this research. Following we describe two more elements we had to integrate in our methodology to ensure the accuracy of the results obtained in the paper.

First, we had to define which pairs of posts should be compared together. A straightforward solution had been to compare, for a given user, each post in  FB to all posts in TW and all posts in G+. However, that option would be inaccurate because we have observed that some users utilize repetitive patterns over time. For instance, we found a user that publishes frequently  posts with the content ``love you my fans", thus following an all to all comparison approach would lead to a wrong classification for quite a lot cross-posts. In order to be accurate and efficient we applied the following methodology. Given a post $P_{FB}$ published by a user $U$ in her FB account at the timestamp $t_{FB}$, we compare $P_{FB}$ with all the posts that user $U$ published in her TW and G+ accounts in a time window starting one week before and finishing one week after $t_{FB}$. In other words, we compare each post in a time window of two weeks around the date that post was published.

Second, TW API limits the number of retrieved posts for any user to the last 3,200 posts she published, while FB and G+ do not have that limitation and provide all the posts published by the user since she registered in the system. Hence, it may happen that for a given user we only have 6 months of posts for TW, but several years for FB and G+. Therefore, in this case it only makes sense to analyze that user for the last 6 months because we would not be able to determine if the information associated to a post published in FB or G+ one year ago was also available in TW at that time. Hence, in order to perform an accurate study, we have restricted our cross-post analysis to the time window imposed by the limitation of TW API for each user in our dataset. It must be noted that the number of posts depicted in table \ref{tab:dataset_info} already consider this limitation.

\emph{We applied the described methodology to the selected 616 OSN professional users and we found 176K cross-posts across their OSNs accounts.}

\begin{table}[t]
  \centering
  \scriptsize
  \caption{Methodology validation, false positive (FP) and false negative (FN) rates of different similarity threshold (ST) in our cross-posting identification methodology.}
    \vspace{-2.5mm}
    \begin{tabular}{c|c|c|c|c|c}
    \hline
    \multicolumn{2}{c|}{ST$>$0.3 similarity} & \multicolumn{2}{c|}{ST$>$0.5 similarity} & \multicolumn{2}{c}{ST$>$0.7 similarity} \\
    \hline
    FP   & FN   & FP   & FN   & FP   & FN \\
    \hline
    15.006 & 0.194 & 0.140 & 1.117 & 0.016 & 4.593 \\
    \hline
    \end{tabular}%
  \label{tab:FP_FN_validation}%
  \vspace{-0.5cm}
\end{table}%

\subsection{Methodology Validation}
In order to ensure the accuracy of the proposed methodology three persons manually classified 12.8K random posts as cross-posts or non-cross-posts.  In order to have a meaningful validation set we ensured that half of the posts had been labeled as cross-post and half as non-cross-posts by our classification tool. Then, given two posts published by a user in two different OSNs we classify them as a cross-post if at least two out of the three persons performing the manual inspection indicate that both posts contain the same information. This allows us to obtain a ground truth set to determine the false positive and false negative rate of our methodology. A false positive occurs when our tool classifies as cross-post two posts (published by the same user in two different OSNs) that are actually referring to a different piece of information. A false negative happens when our tool classifies as non-cross-post two posts that actually contains the same information.

Based on the ground truth set we compute the false negative and false positive rate for our methodology using three different thresholds for the second step of the algorithm: $0.3$, $0.5$ and $0.7$. Basically, a lower threshold requires less similarity between the compared posts to classify them as cross-post. Table \ref{tab:FP_FN_validation} shows the false positive and false negative rate for our algorithm when it uses each of the evaluated thresholds. The results clearly determine that 0.5 is a very good threshold since it presents a very low rate for false positives (0.14\%) and false negatives (1.11\%). However, on the one hand, a threshold of 0.3 imposes a very low similarity to classify two posts as cross-post and thus it presents an unacceptable false positive rate (15\%). On the other hand, a threshold of 0.7 is too strict and it skips some pairs that actually contains the same information and classify them as non-cross-posts, thus presenting a poor false negative rate (4.5\%). Therefore, based on this experimental validation of our methodology, we decided to establish a threshold of 0.5 in our algorithm.

\section{Cross-Posting Characterization}
\label{sec:cross_volume}

The first question we aim to answer in this section is whether the cross-posting phenomenon exist in the activity of professional users, and what is its weight in FB, TW and G+. Following, we look at how this cross-posting occurs among the three OSNs under analysis. To this end, we quantify what is the volume of cross-posting happening between FB-G+, FB-TW, TW-G+ and FB-TW-G+, in order to determine what pair of OSNs is actually sharing more common information. Finally, we also want to characterize the impact of cross-posting in the attracted engagement measured in terms of likes comments, and shares.

\begin{figure}[t]
	\centering
	\includegraphics[width=0.3\textwidth,height=0.26\textwidth]{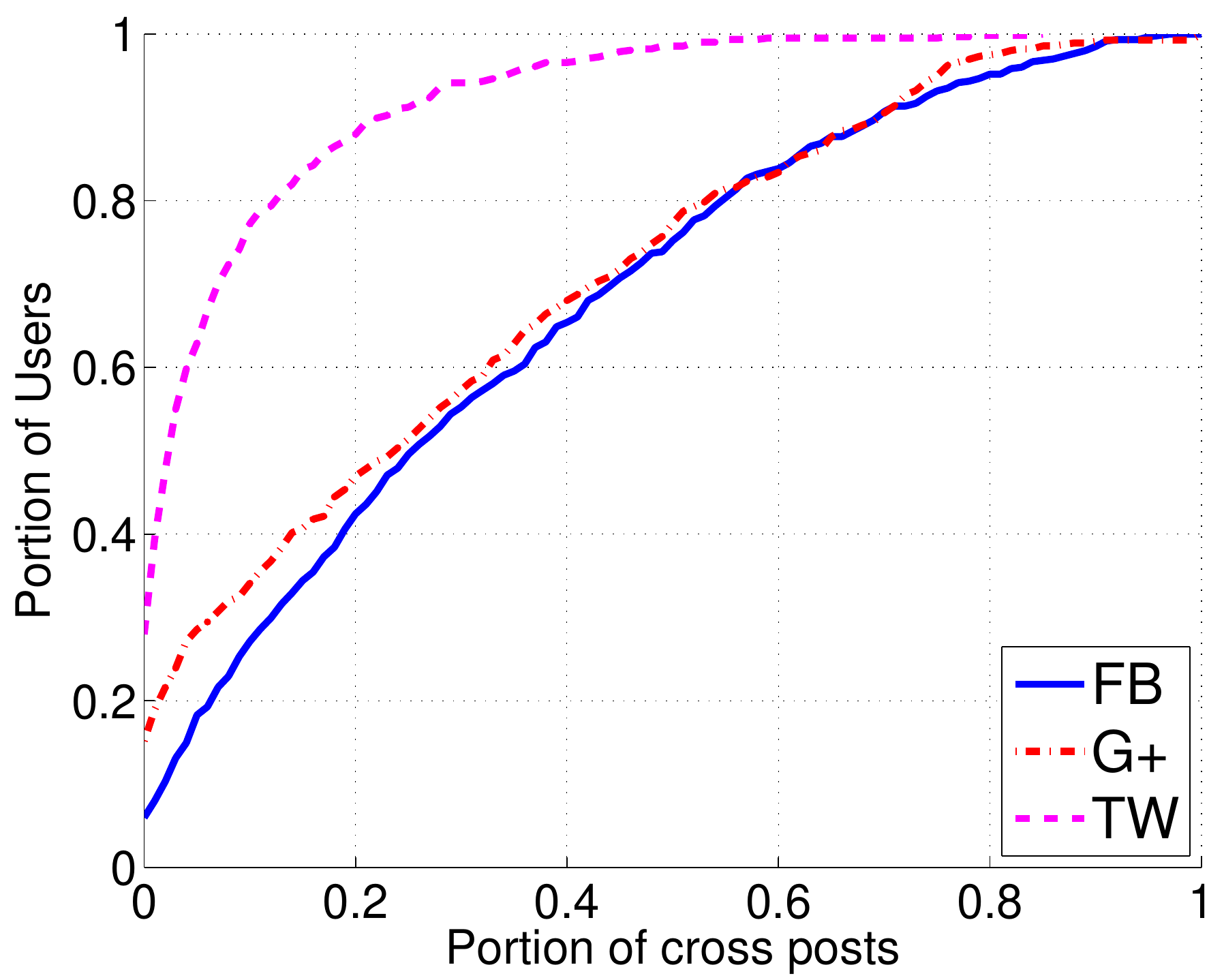}
	\vspace{-0.5cm}
	\caption{CDF for portion of cross-posts per user in FB, G+ and TW}
	\label{fig:perc_total_cross_posts}
\vspace{-0.4cm}
\end{figure}

%

\begin{table}[t]
	\centering
	\scriptsize
	\caption{median and average values for absolute number (and percentage) of Cross posts across users in FB, G+ and TW}
\vspace{-0.25cm}
		\begin{tabular}{c|cc|cc}
    \hline
		         & \multicolumn{2}{c}{\#Cross Posts} & \multicolumn{2}{|c}{Portion of cross posts (\%)} \\
    \hline
			OSNs & Median & Average & Median & Average \\
    \hline
    FB   & 114  & 231  & 26.42 & 32.14 \\
    G+   & 20   & 83   & 24.50 & 29.31 \\
    TW   & 85   & 196  & 3.34 & 7.36 \\
    \hline
		\end{tabular}%
\vspace{-0.35cm}
		\label{tab:median_mean_Cross_posts}%
\end{table}%



\subsection{Quantification of cross-posting activity}

\begin{figure*}[t]
\begin{minipage}[b]{0.3\linewidth}
	\centering
	\includegraphics[width=0.74\textwidth,height=0.7\textwidth]{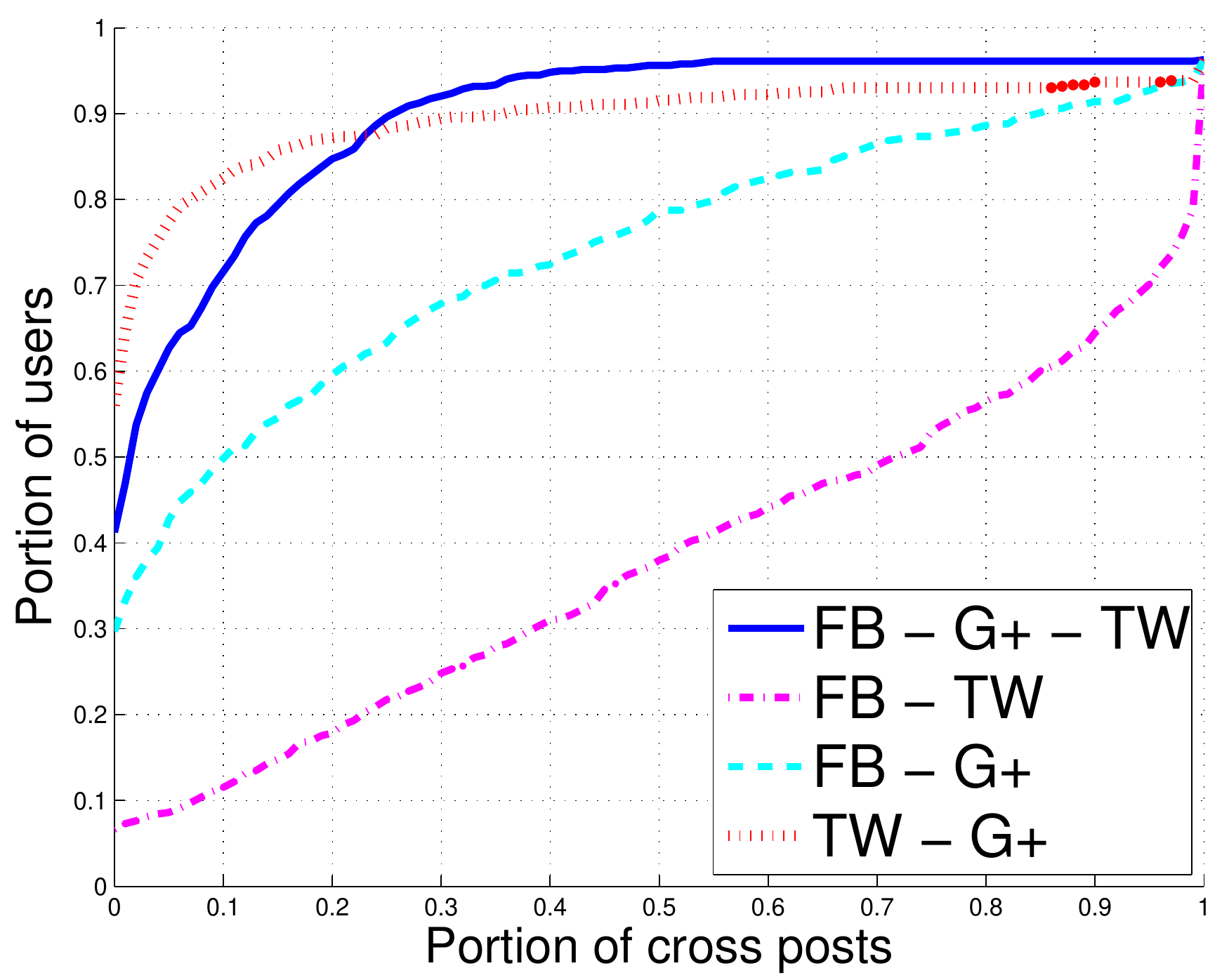}

	\vspace{0.15cm}
	\caption{CDF for portion of cross posts in each possible cross-posting pattern across users.}
	\label{fig:perc_identified_cross_posts}
	\vspace{-0.2cm}
\end{minipage}
\hspace{0.2cm}
\begin{minipage}[b]{0.68\linewidth}
	\centering
	\subfigure[{FB}]{\includegraphics[width=0.31\textwidth,height=0.31\textwidth]{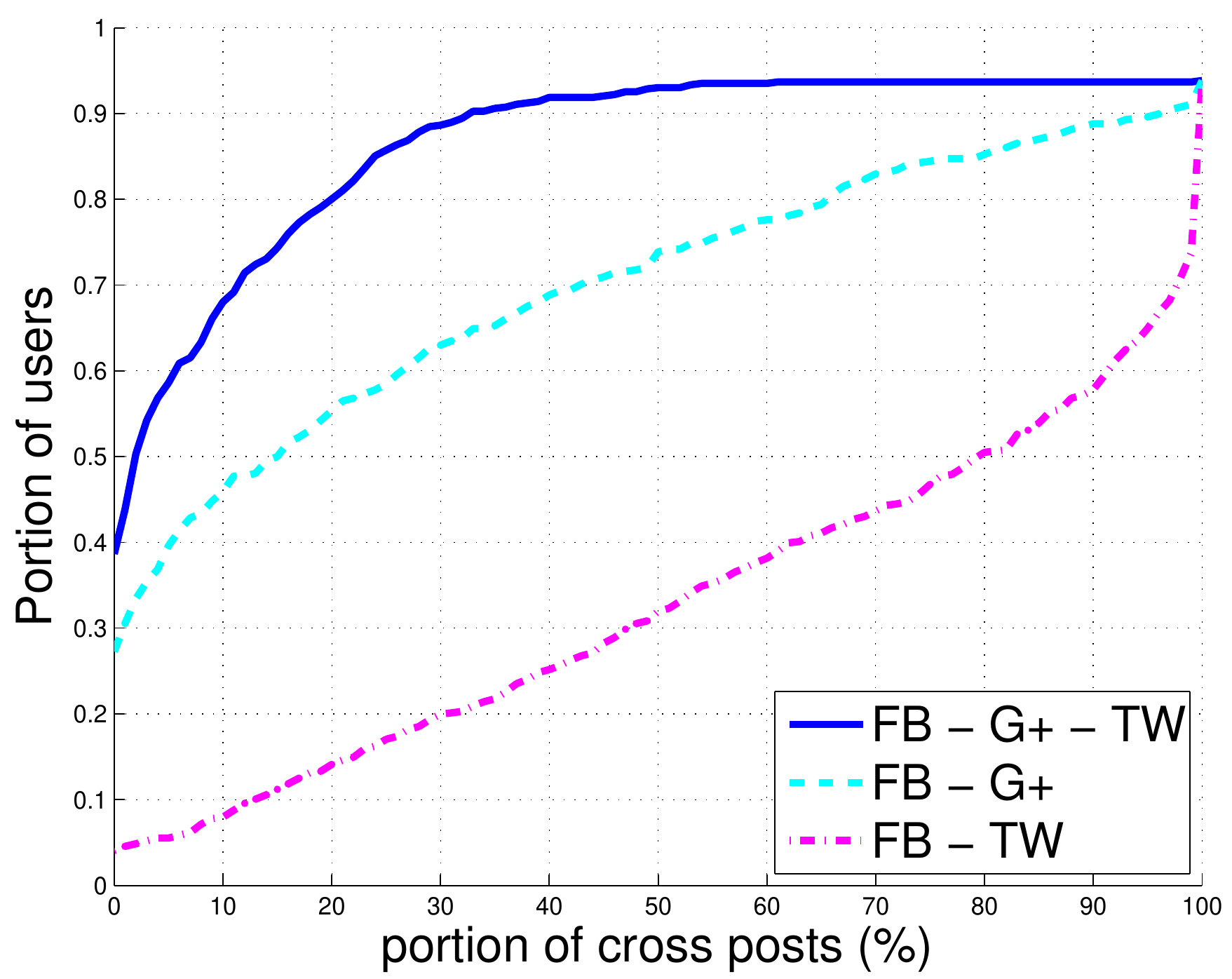}	

	\label{fig:perc_identified_posts_per_osn_FB}}
	\subfigure[{G+}]{\includegraphics[width=0.31\textwidth,height=0.31\textwidth]{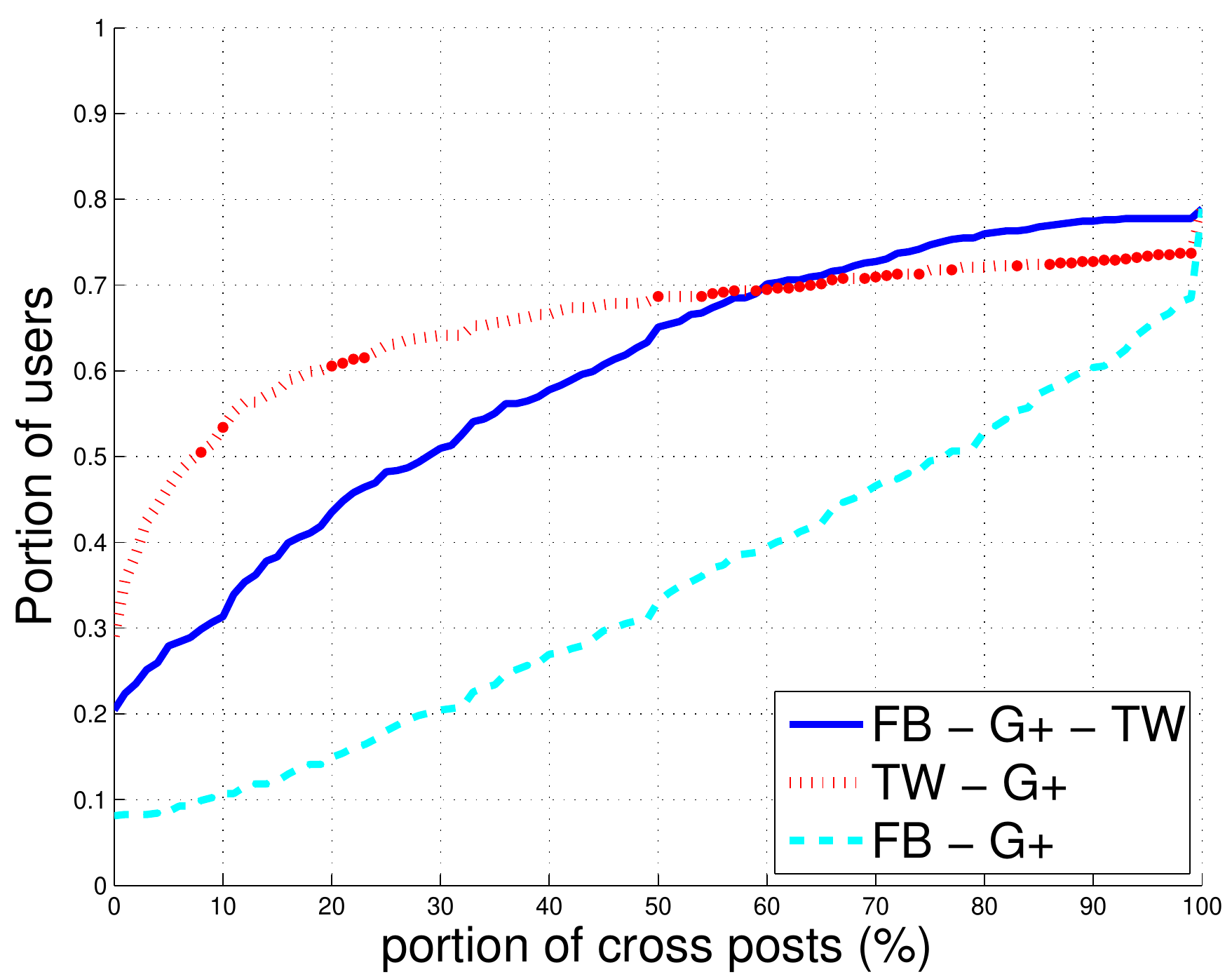}

	\label{fig:perc_identified_posts_per_osn_G+}}	
	\subfigure[{TW}]{\includegraphics[width=0.31\textwidth,height=0.31\textwidth]{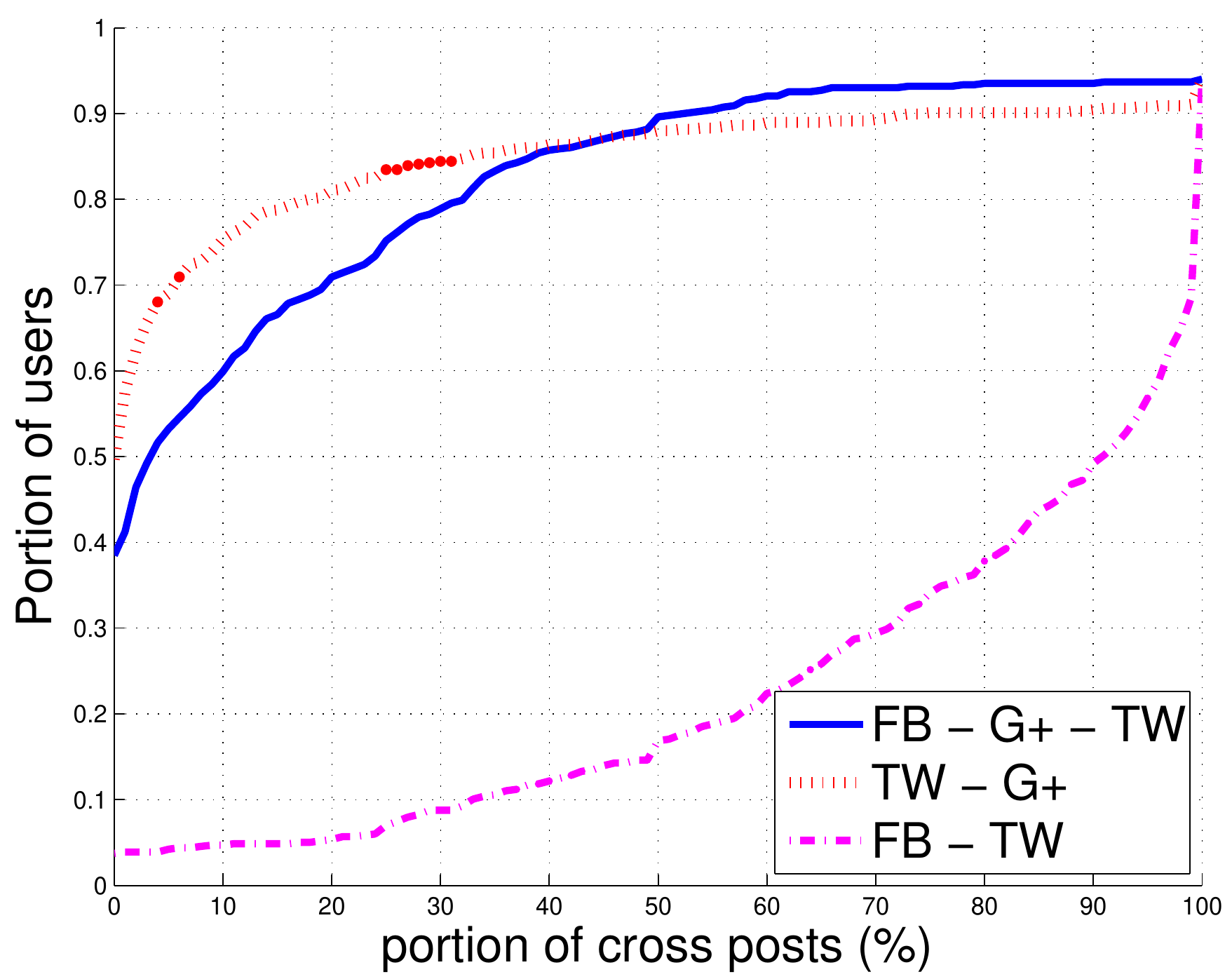}
	
	\label{fig:perc_identified_posts_per_osn_TW}}
	\vspace{-0.2cm}	
	\caption{CDF for portion of cross posts in each possible cross-posting pattern for the cross-posts initiated in FB (a), G+ (b) and TW (c)}
	\label{fig:perc_identified_posts_per_osn}
\vspace{-0.2cm}
\end{minipage}
\end{figure*}

The goal is to quantify the cross-posting phenomenon for professional users in FB, TW and G+.  Towards this end, we compute for each user and each OSN the portion of cross posts with respect to all the posts each user has published. For instance, given a user $U$ and her FB account we compute how many posts published in that account also appear in TW, G+ or both. We quantify the same parameter for the TW and G+ accounts of user $U$\footnote{ It must be noted that for this analysis we do not take into account where the post appears first, but only consider whether it is unique in an OSN or it appears in 2 or 3 of them.}.

Figure \ref{fig:perc_total_cross_posts} shows the CDF for the portion of cross posts across the 616 users analyzed in the three OSNs. The x axis refers to the portion of posts and the y axis to the portion of users. For instance, the point \{x=0.2, y=0.4\} in the line associated to FB indicates that 40\% of the users have $\leq20\%$ of cross-posts in their FB accounts.

The first immediate conclusion extracted from the graph is that most of the professional users have published some cross-post. In particular, when we consider FB accounts we find that only 6\% of the users do not have any cross-post, which means for those users the information published in FB can be found in neither in TW nor in G+. This number grows up to 15\% and 28\% for G+ and TW, respectively. Therefore, a vast majority of professional users published some cross-post at some point. Hence, the first conclusion is that in general professional users find some value on the cross-posting activity.

If we compare the results obtained for the three OSNs, we clearly observe that, in relative terms, the cross-posting activity is more frequent for those posts published in FB and G+ than in TW. The results for TW show that  most of the tweets are not replicated neither in FB nor in G+. The median value, which shows the typical portion of cross-posts for a user in each OSN, shows that for a typical professional user around 1/4 of the posts that appear in FB and 1/4 of the posts that appear in G+ are also available in at least one more OSN. However, in the case of TW,  out of 100 tweets only 3 of them are replicated in other OSNs. Finally, we can find quite a lot professional users with an intensive cross posting activity. In particular, 25\%, 23\% and 1.5\% of the analyzed users, in FB, G+ and TW, respectively, have published more cross posts (i.e., $\geq 50\%$) than posts appearing exclusively in a single OSN. We refer to these posts as \textit{non-cross-posts}.

The previous analysis refers to the cross-posting activity in relative terms. However, it is important to notice that, according to the overall activity of the professional users in our dataset, the publishing rate of professional users in TW is 4$\times$ higher than in FB and G+. Table \ref{tab:median_mean_Cross_posts} presents the median and average values for the absolute number and portion of cross-posts per user in each OSN. The results reveal that although TW presents a much lower cross-posting activity in relative terms, it actually has a larger number of cross-posts than G+, and it is much closer to FB in absolute cross-posts number.

%

\begin{figure*}[t]
	\centering
	\subfigure[{Likes}]{\includegraphics[width=0.3\textwidth,height=0.3\textwidth]{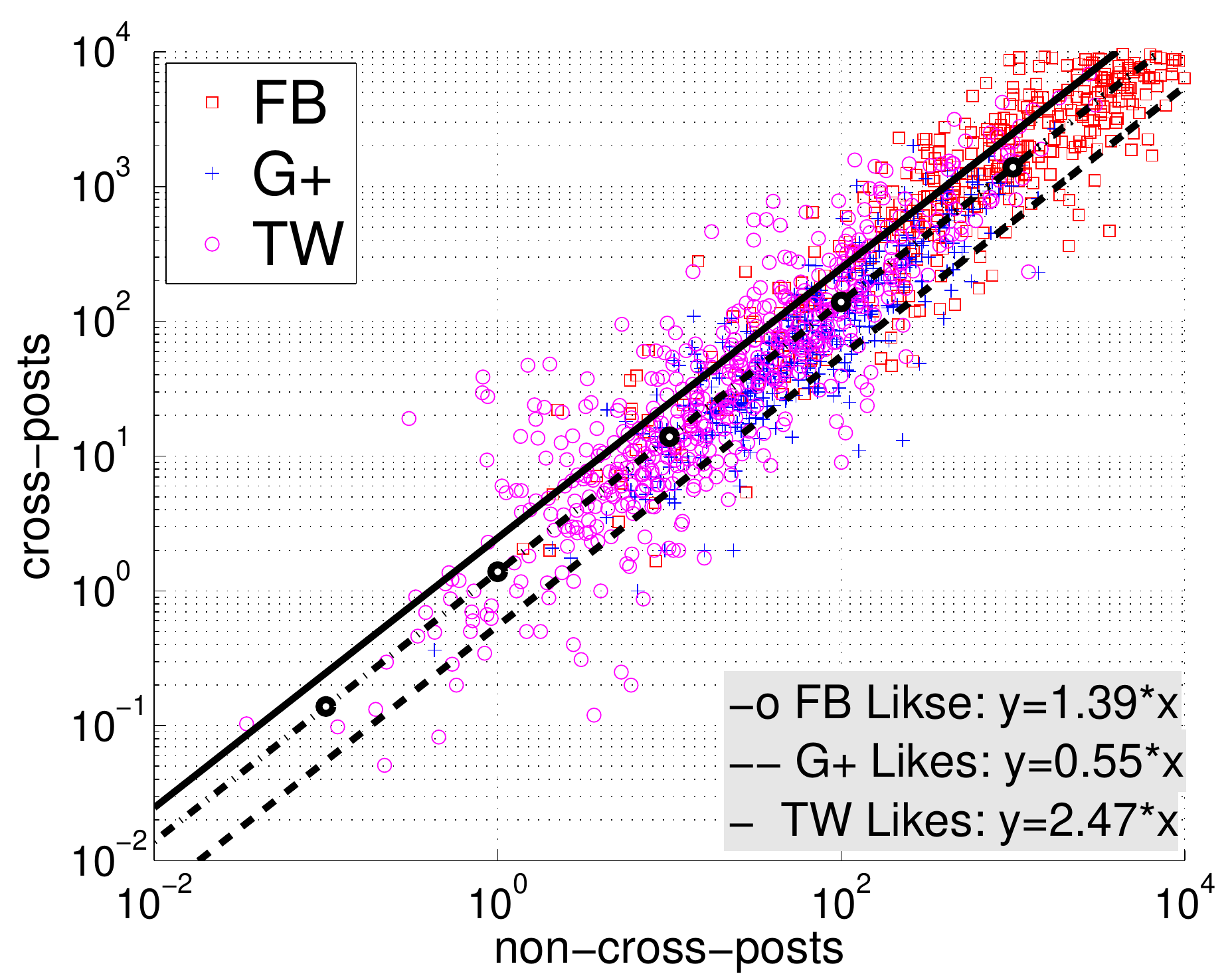} 	
	
	\label{fig:All_likes}} \subfigure[{Comments}]{\includegraphics[width=0.3\textwidth,height=0.3\textwidth]{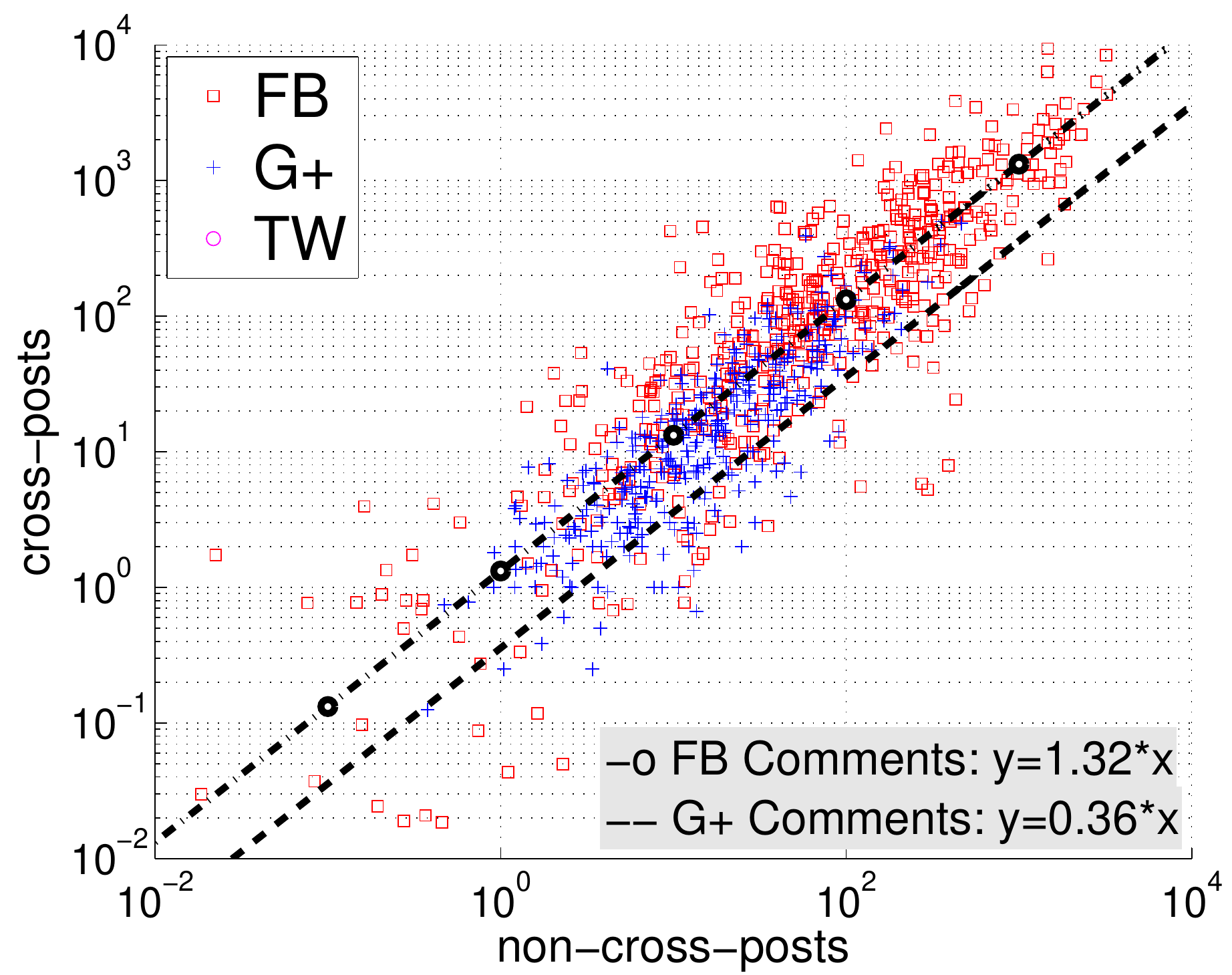}
	\label{fig:All_comments}}
	\subfigure[{Shares}]{\includegraphics[width=0.3\textwidth,height=0.3\textwidth]{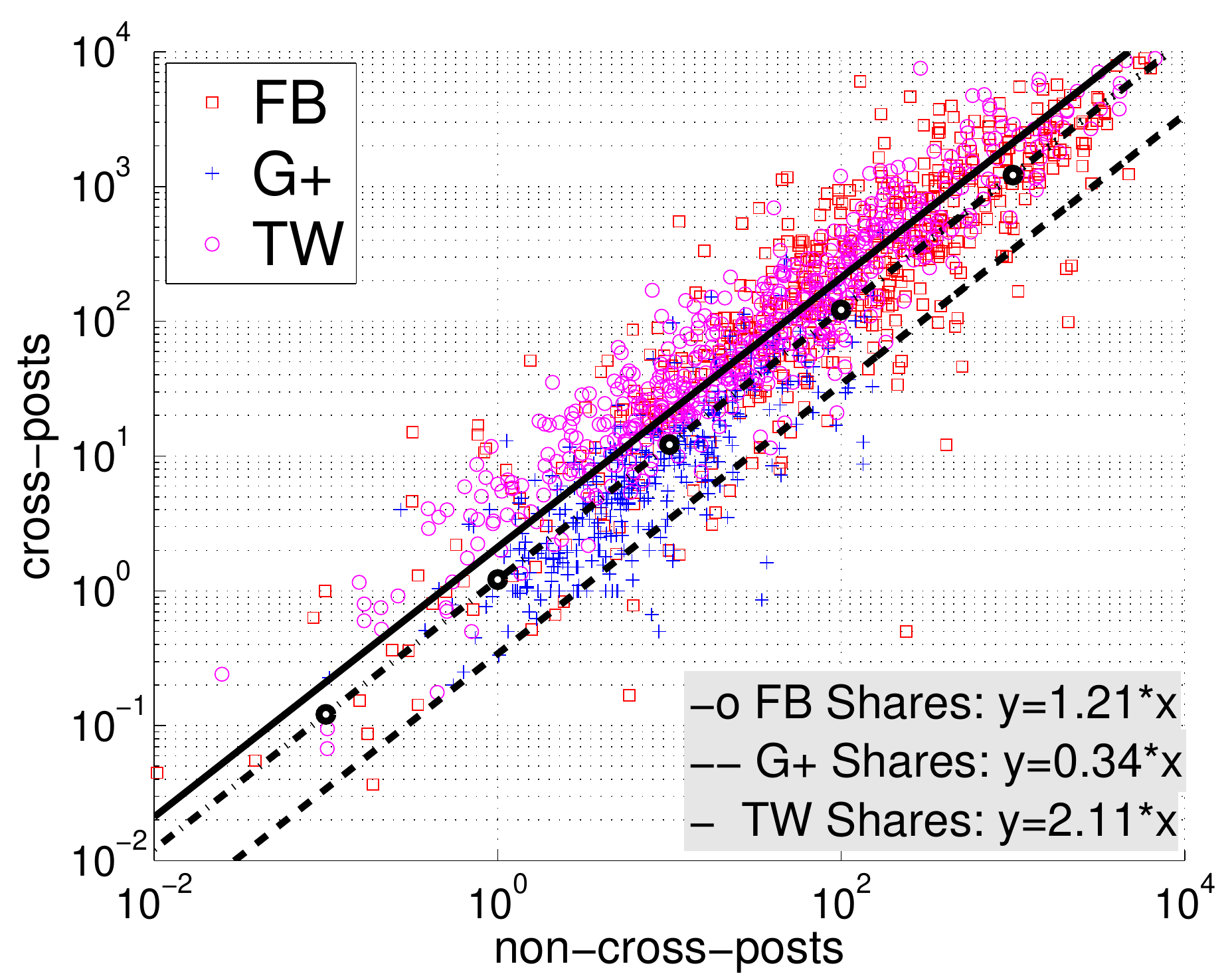}	
	\label{fig:All_shares}}
	\vspace{-0.35cm}
	\caption{Users' average attracted engagement per post, for cross posts initiated in each OSNs vs. non-cross posts.}
	\label{fig:Croos_nonCross_original}
\vspace{-0.5cm}
\end{figure*}

\subsection{Inter-OSN Cross-Posting}
Once we have demonstrated that cross-posting is a common practice among professional users in FB, TW and G+, we analyze how cross-posting happens among them. Then, our goal is to find whether professional users prefer to share things in FB and TW, or rather it is more frequent finding common posts in FB and G+, or actually there are lots of cross-posts published in TW and G+.

In order to perform this analysis we proceed as follows. For a given user $U$ we get all her cross-posts in FB (independently whether the first appearance happened in that OSN or another one) and compute which portion of them also appears in TW, which portion in G+ and which portion in both TW and G+. We repeat the same process for the TW and G+ accounts of user $U$. Therefore, for each user we know what is the cross-posting level for the following relations: \textsl{FB-TW}, \textsl{FB-G+}, \textsl{TW-G+} and \textsl{FB-TW-G+}.

Figure \ref{fig:perc_identified_cross_posts} shows the CDF for the portion of cross-posts that occurs for the four referred relations across the users in our dataset. Again in this figure the x axis refers to portion of posts and the y axis shows the portion of users. Then for instance the point \{x=0.4, y=0.3\} in the \textsl{FB-TW} line indicates that 30\% of the users publish $\leq$40\% of their cross posts in FB and TW.

The results in the figure demonstrate that professional users perform much more cross-posting  between FB and TW than in any other combination of OSNs.  This claim is supported by the fact that in median a professional user publishes 70\% of their cross-posts in FB and TW. In addition, we can only find 8\% of the users that never shared a post between their FB and TW accounts, while this value grows to 30\% between FB and G+, to 40\% for the case in which the three OSNs are involved, and to 55\% when we consider TW and G+. Therefore, this last result surprisingly states that is more likely that a user publishes the same posts in the three OSNs than only in TW and G+.

In order to complete this analysis we repeat the experiment for each OSN by only considering those cross-posts that were first published in each OSN. Figures \ref{fig:perc_identified_posts_per_osn_FB}, \ref{fig:perc_identified_posts_per_osn_G+} and \ref{fig:perc_identified_posts_per_osn_TW} present the results for the cross-posts originated in FB, TW and G+, respectively.

For the case of FB, we observe that a typical user (based on the results around the 45th and 55th percentile) roughly follows the next pattern for the cross-posts that are first published in FB: 75-85\% are replicated only in TW, 10-20\% only in G+ and around 5\% in both TW and G+.  In addition, in this case less than 5\% of the users do not share any cross-post started in FB with TW, while this value reaches 30\% and 40\% for the case of G+, and the case when a post started in FB is replicated in both TW and G+, respectively. In the other extreme, we can find 40\% of the users that only use TW for 90\% of their cross-posts initiated in FB. Finally, we could not find any user that only chose TW and/or FB to republish all the cross-posts that appeared first in FB.

In the case of G+, a typical user republishes exclusively in FB around 70-75\% of the posts initiated in G+, 25-30\% in both FB and TW, and only 5-10\% in TW. This result reveals that users believe that the information published first in G+ is definitely more interesting to FB audience. In addition, the results reinforces the idea that users do not see much interest on cross-posting information between G+ and TW. Finally, it is interesting to notice that none of the users in our dataset has a flow higher than 80\% of the cross-posts initiated in G+ for any of the possible combinations.

For TW we find the most extreme case. A typical TW user republishes 90\% of her cross-posts initiated in TW exclusively in FB, while she rarely republish in both in FB and G+ (around 5\%) and even more rarely only in TW.  These results confirm that also in the direction $TW -> G+$ there is very few cross-posting.

In a nutshell, we can find more cross-posting between FB and TW (in either direction) than with G+, while the specific cross-posting between  TW and G+ (in either direction) appears as the least preferred option, since users prefer to publish the information in all the three OSNs than only in TW and G+.

\subsection{Engagement Analysis}

\begin{table*}[t]
	\centering
	\scriptsize
	\caption{Portion of cross-posts published for first time in FB, TW or G+ for different cross-posting patterns: FB-TW-G+, FB-TW. FB-G+, TW-G+. The table also includes the portion of posts that are published in at least two OSNs at the same time (i.e., exact timestamps)}
\vspace{-0.25cm}
	\begin{tabular}{c|rc|ccc|c}
    \hline
   cross-posting pattern   & \#Posts & \%Posts & \%FB (1st) & \%G+ (1st) & \%TW (1st) & \%posts with same publishing time \\
    \hline
    FB - G+ - TW & 18619 & 10.56 & 34.93 & 12.32 & 49.80 & 2.95 \\
    FB - G+ & 34337 & 19.48 & 73.68 & 24.07 & - & 2.26 \\
    FB - TW & 117276 & 66.52 & 36.28 & - & 56.80 & 6.92 \\
    G+ - TW & 6073 & 3.44 & - & 23.79 & 75.96 & 0.25 \\
    \hline
	\end{tabular}%
	\label{tab:perc_of_preferences}%
\vspace{-0.3cm}
\end{table*}%

A plausible reason of why professional OSN users publish the same information across different OSNs is to try to increase the coverage in order to engage as many end-user as possible within their accounts. Therefore, in this subsection we want to conclude whether cross-posts achieve more engagement than non-cross-posts in FB, TW and G+. In order to measure the engagement we use standard reaction mechanisms available for end users in OSNs: likes, comments and shares\footnote{This is the nomenclature employed in FB. A like is associated to a +1 in G+ and to a favourite in TW. A share is associated to reshare in G+ and a retweet in TW.}. As we acknowledge in Section \ref{sec:methodology}, our TW data collection tool could not retrieve comments. These reaction mechanisms allow professional users to interact with end-users through its OSN account and obtain a very valuable first-hand feedback from them. Therefore, engaging as many end-users as possible is an important goal for professional OSNs users.

In order to measure the efficiency of cross-posts to attract engagement in one OSN we measure, for a given user $U$, the average engagement for $U's$ non-cross-posts and  $U's$ cross-posts initiated in that OSN in terms of likes, comments and shares. We apply this methodology to all the users for their FB, G+ and TW accounts.
Figure \ref{fig:Croos_nonCross_original} shows a scatter plot for FB, G+ and TW for each of the engagement type: likes (Figure \ref{fig:All_likes}), comments (Figure \ref{fig:All_comments}) and shares (Figure \ref{fig:All_shares}). Each point in the graphs represents a user with an x coordinate referring to the average engagement for non-cross-posts and y coordinate referring to the average engagement for cross-posts initiated by that user in that OSN. In addition, all the figures include three lines (one per OSN) showing the linear regression for the cloud of points represented by an equation\footnote{Usually a linear regression is represented as y=ax+b, but in the figure we just use y=ax, since we are interested in the slope, but not in the offset} of type $y=ax$. When the slope of the linear regression, represented by the value of $a$, is greater than 1, it means that for that OSN cross-posting is worthy since cross-posts attract more engagement than non-cross-posts in average.

The results demonstrate that cross-posts in FB and TW allows professional users to attract more engagement than non-cross-posts. However in the case of G+ cross-posts receive considerably less attention than non-cross-posts. In more detail, a FB user attracts 39\% more likes, 32\% more comments and 21\% more shares in FB when she uses cross-posts instead of non-cross-posts.
In the case of TW cross-posting provides even more benefit in term of engagement. This is, a cross-post initiated in the TW account of a professional user attracts 2.47$\times$ and 2.1$\times$ more likes (i.e., favourites) and shares (i.e., retweets) than a non-cross-posts. Finally, in the case of G+ a cross-post roughly achieves 1/2 of the likes (i.e, +1's), 1/3 of the comments and 1/3 of the shares compared to non-cross-posts. Therefore, cross-posting seems to be a bad strategy if the goal of a professional user is to attract as many reactions as possible in G+.
Finally, it should be mentioned that the presented result in this section does not imply any causal relationships between cross-posting and engagement increment.

\emph{In summary, cross-posting exists and it is a frequent practice across professional users in FB, TW and G+. It mostly happens between the FB and TW accounts of professional users, and it very rarely occurs between TW and G+. Finally, in terms of attracted engagement, cross-posting is beneficial in FB and TW, but not in G+.}

\section{Preference of Professional Publishers}
\label{sec:preferece}

Professional users utilize OSNs to interact with their followers and share with them more or less relevant information. In previous section we have demonstrated that quite frequently an end-user can find the same information in two (or more) OSNs. Based on this finding, in this section we tackle two interesting questions. First, we want to know in overall which OSN is used more frequently as first option to publish fresh information that later will be republished in other OSNs. Second, we want to understand what is the OSN that professional users prefer to publish first the information.
Answering the first question will determine which OSN is used more times as source of cross-OSN information, while the response to the second question will roughly determine what is the OSN that professional users value more to publish first their fresh updates.

\begin{table}[t]

\begin{minipage}[b]{0.58\linewidth}
  \centering
  \vspace{-0.15cm}
  \scriptsize
	\caption{Cross-Posts initiated in FB, G+, TW}
  \vspace{-0.25cm}
    \begin{tabular}{c|rc}
    \hline
    OSN  & \#Posts & \%Posts \\
    \hline
    FB   & 74355 & 42.17 \\
    G+   & 12002 & 6.81 \\
    TW   & 80497 & 45.66 \\
    other & 9451 & 5.36 \\
    \hline
    \end{tabular}%
  \label{tab:prefered_OSN_per_Post}%
\end{minipage}
\begin{minipage}[b]{0.4\linewidth}
	\centering
	\scriptsize
	\caption{Preferred OSN per user}
  \vspace{-0.2cm}
	\begin{tabular}{c|rc}
    \hline
    OSN  & \#Users  & \%Users \\
    \hline
    FB   & 307  & 50 \\
    G+   & 30   & 5 \\
    TW   & 275  & 45 \\
    \hline
	\end{tabular}%
	\label{tab:Preferred_OSN}%
\vspace{-0.5cm}
\end{minipage}
  \vspace{-0.4cm}
\end{table}%


\subsection{OSN-based Analysis}
Table \ref{tab:prefered_OSN_per_Post} shows the number and portion of cross-posts in our dataset that were initiated in FB, TW and G+. The results demonstrate that TW appears as  initial source of information for 45\% of the cross-posts closely followed by FB with 42\%, while G+ is rarely chosen as first option. Finally, we find a very interesting result associated to the category \textit{``other"} that represents those cross-posts that could not be assigned to a particular OSN since they were published exactly at the same time (i.e., same timestamp) in at least two OSNs. It is surprising that almost 10K cross-posts, which represent 5.3\% of all the cross-posts in our dataset, experienced this parallel publication. This reflects the use of automatic publishing tools that upload in parallel some information to two or more OSNs.

As we determined in the previous section, most of the posts are not published in all the three OSNs, but just two of them. Therefore, it is interesting to analyze for each particular publishing pattern which OSN appears more frequently as initial source of information. Table \ref{tab:perc_of_preferences} shows the results for all the possible cross-post patterns: \textsl{FB-TW-G+}, \textsl{FB-TW,} \textsl{FB-G+} and \textsl{TW-G+}. First of all, the results confirm the conclusion obtained in the previous section since 2/3 of the cross-posts appear exclusively in FB and TW, 1/5 belong to the category \textsl{FB-G+}, and as we already stated it is more likely finding cross-posts across the three OSNs (10\%) than only across G+ and TW (3.4\%). In the most popular category, i.e., \textsl{FB-TW}, TW appears as first option for 57\% of the posts while FB is chosen in first place only 36\% of the times. When G+ competes individually either with FB or TW, it is source of information only 1/4 of the times.  For those posts published in the three OSNs, 1/2 of them appear first in TW, 1/3 in FB and 1/10 in G+.

Finally, we want to highlight that all the categories include some portion of posts that where published in parallel at the same exact time in two OSNs. This phenomenon is especially relevant for cross-posts between \textsl{FB-TW}.

In summary, the OSN-based analysis demonstrates that Twitter is the OSN selected as initial source of information more frequently. FB appears as the second option close to Twitter. Finally,  G+ is the least preferred option.




\subsection{User-based Analysis}

The OSN-based analysis revealed that Twitter is chosen as first option for a larger number of cross-posts. However, we cannot extract from that analysis that TW is the preferred OSN for most of the users, since it may happen that very active users contributing a large number of posts prefer TW but less active users prefer FB or G+. Therefore, in this section we analyze which is the preferred OSN for professional users. For a given user its preferred OSN is the one she selected in first place for a major number of posts. For instance, if a user has generated 20 cross-posts from which 10 were first published in FB, 6 in G+ and 4 in TW, we define FB as the preferred OSN for that user. Table \ref{tab:Preferred_OSN} shows the number and portion of users in our dataset that prefer each OSN. The results reveal that half of the professional users prefer FB, closely followed by 45\% of the users that prefer TW, while only 5\% of the users chooses G+ as initial OSN for publishing their post. Therefore, FB and TW has exchanged their positions as compared to the OSN-based results. As we indicated above, the difference between the post-based and user-based results comes from the fact that users tend to be more active in TW. 

Once we have classified professional users' preference, a subsequent question is, can we find users that shows a strong preference for a particular OSN? In other words, are there users that utilize as source of information one single OSN for most of their cross-posts?

Table \ref{tab:user_preference_} shows the number and portion of professional users in our dataset that choose either FB, TW or G+ to initiate 100\% or 80\% of their cross-posts showing a clear strong preference. In addition, we also quantify the number and portion of users that publish in first place less than 50\% of their posts in all three OSNs and thus do not show any strong preference.
We can find 19, 11 and 2 users that always choose TW, FB and G+ as initial source for their cross-posting activity, respectively. If we move down the threshold to 80\% the number of users showing a clear evidence of which OSN they prefer grows a lot for FB and TW, but not for G+ that only accounts for 5 users. There are 75 (12.18\%) users with a preference for FB and 102 (16.56\%) with a noticeable preference for TW. In contrast to these users showing a clear OSN preference, we can find 95 (15.4\%) users that are not biased towards any OSN, even though they make use of cross-posts.

\begin{table}[t]
  \centering
	\scriptsize
  \vspace{-0.05cm}
  \caption{ Users classification based on different OSN preference criteria: $(i)$ users initiating 100\% of their cross-posts from one OSN; $(ii)$ users initiating $\geq$80\% of their cross posts from one OSN; $(iii)$ users starting $<$50\% of their posts from all three OSNs.}
    \vspace{-.25cm}
    \begin{tabular}{l|c|cc|cc|cc}
    \hline
    Criteria & \#User & \#FB & \%FB   & \#G+ & \%G+   & \#TW & \%TW \\
    \hline
    100\% & 32   & 11   & 1.79 & 2    & 0.32 & 19   & 3.08 \\
    $\geq$80\% & 182  & 75   & 12.18 & 5    & 0.81 & 102  & 16.56 \\
    $<$50\% & 95 & - & - & -  & -  & -  & - \\
    \hline
    \end{tabular}%
  \label{tab:user_preference_}%
  \vspace{-0.5cm}
\end{table}%

\emph{In summary, professional users are (more or less) equally divided into those that prefer TW and those that prefer FB, and very few cases that show a preference for G+.}

\section{Cross-posting Behavioural Patterns}
\label{sec:behavior}

\begin{figure*}[t]
	\centering
	\subfigure[{CDF of cross-post similarity}]{\includegraphics[width=0.32\textwidth, height=0.25\textwidth]{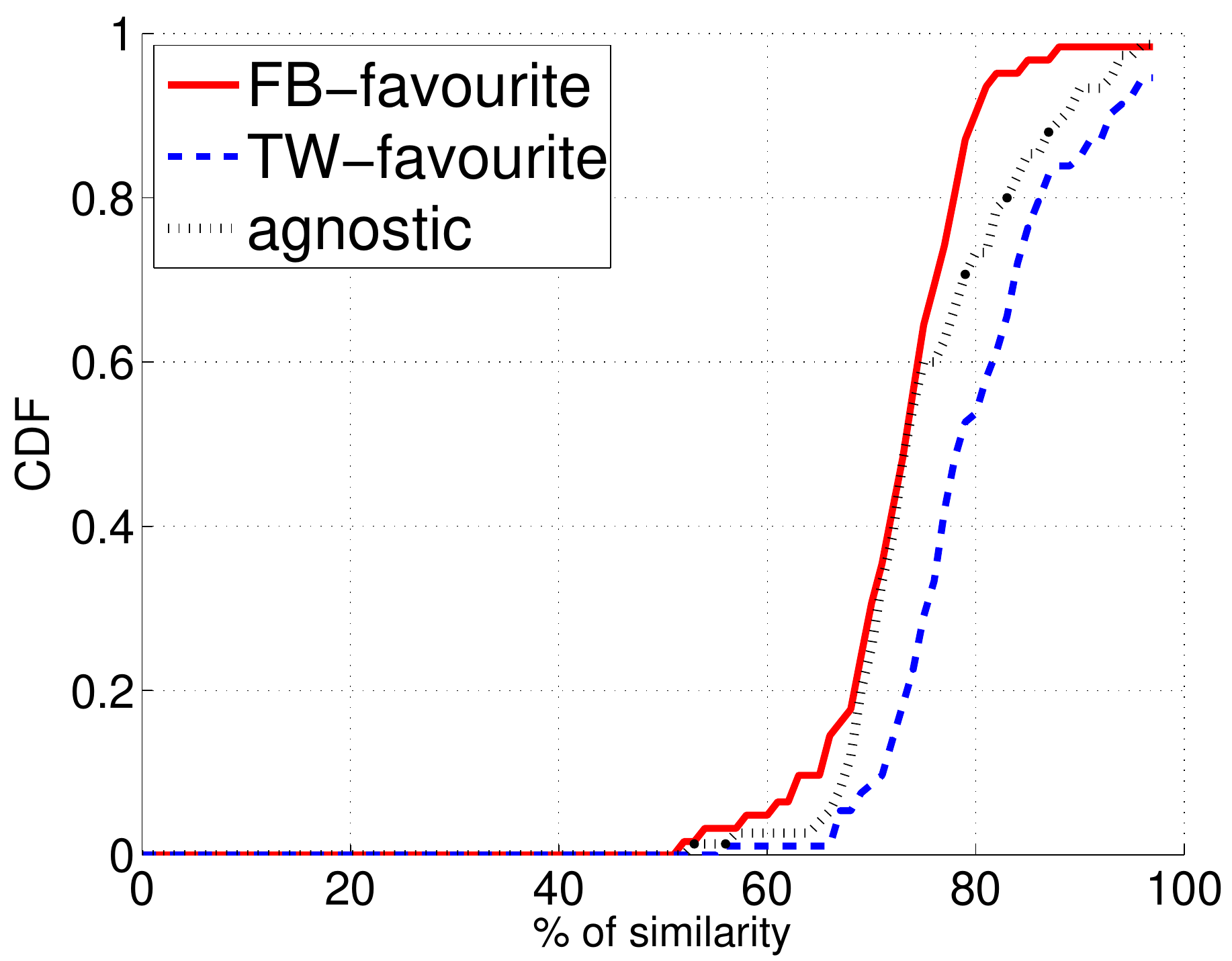}
	\label{similarity_preference_FBTW}}
	\subfigure[{Type of post}]{\includegraphics[width=0.32\textwidth, height=0.26\textwidth]{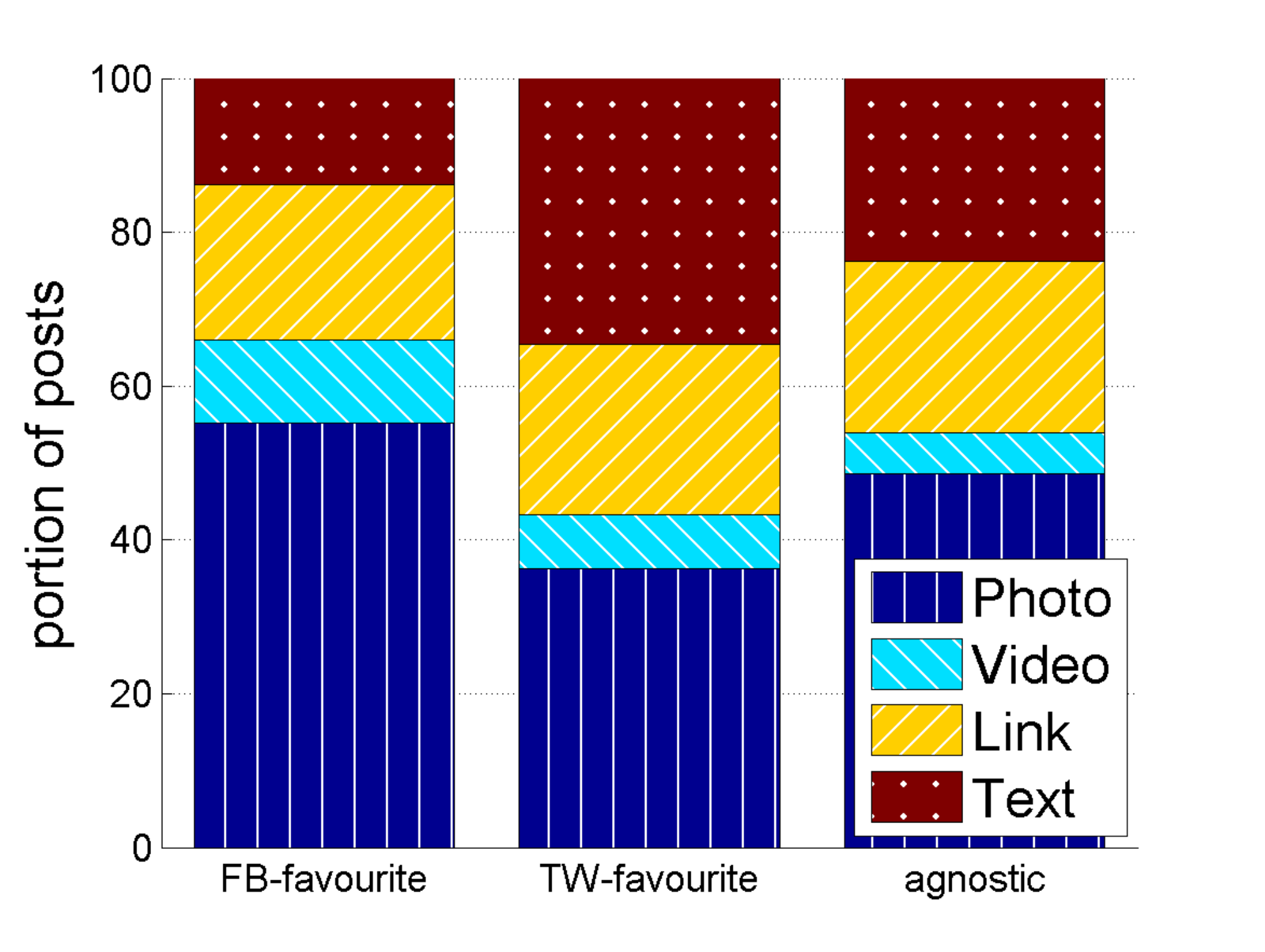}
	\label{post_type_preference_FBTW}}
	\subfigure[{Website associated to urls}]{\includegraphics[width=0.32\textwidth, height=0.26\textwidth]{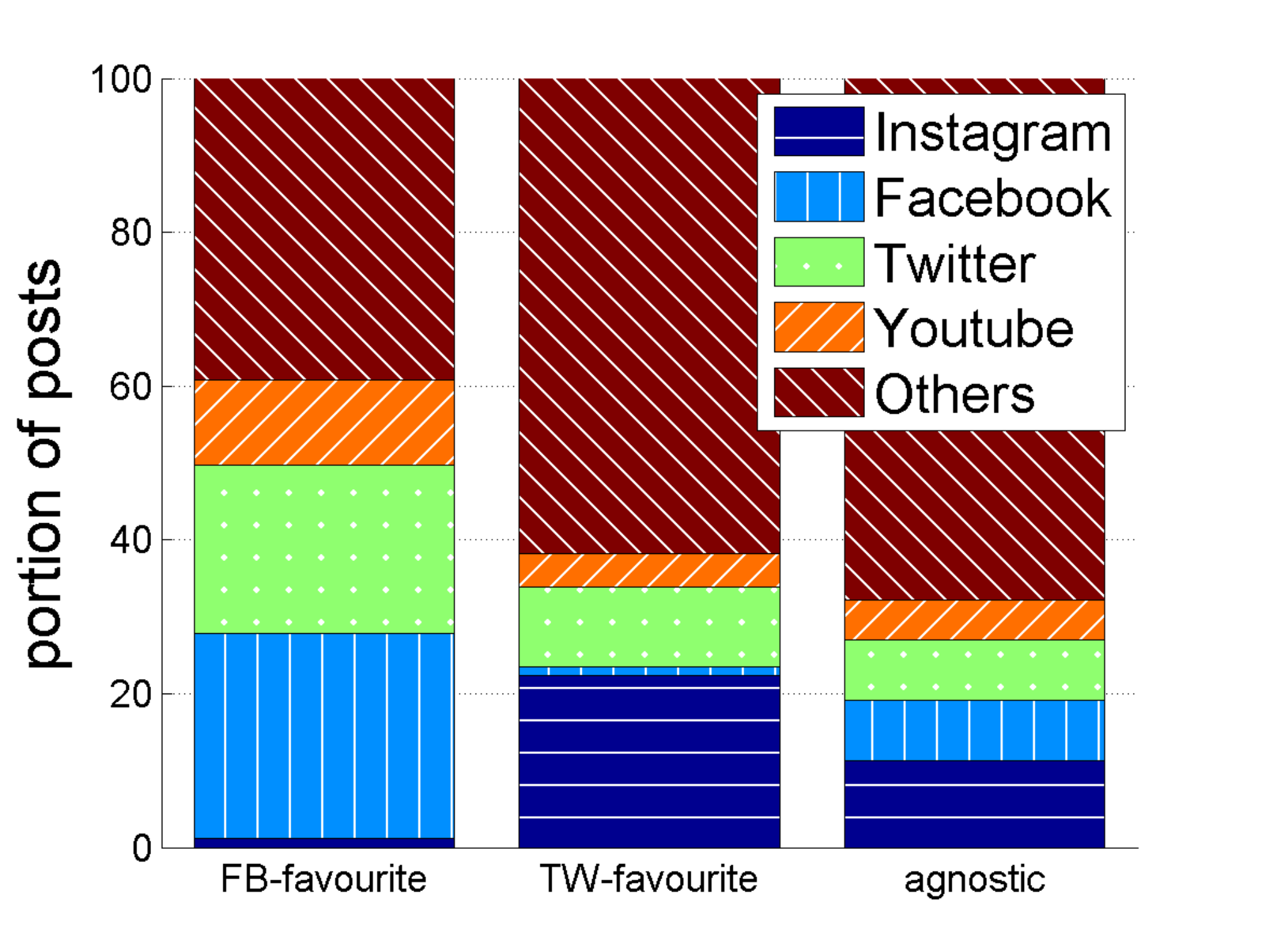}
	\label{short_url_preference_FBTW}}
	\vspace{-0.3cm}
	\caption{Cross-posting behaviour characterization based on professional users preference.}
\vspace{-0.4cm}
\end{figure*}

We have fully characterized the cross-posting phenomenon as well as what is the preferred OSN for professional users in the context of cross-posting. To finalize this paper we want to explore the presence of explicit differences in the cross-posting activity for groups of users presenting different but well defined profiles according to a given characteristic. We will focus on two characteristics: $(i)$ OSN preference and $(ii)$ inter-posting interval. First, the goal is to determine whether there are significant behavioural differences in the cross-posting pattern for professional users showing a strong preference for TW, professional users showing a strong preference for FB and agnostic users. Second, we separately analyze whether professional users publishing their cross-posts in two OSNs in a short time window show some behavioural differences compared to users that delay a lot the publication of the cross-posts in the second OSN.  We refer to the time window between the publication in the first OSN and the second OSN as \textit{inter-posting interval}.

We characterize the cross-posting behaviour using three parameters that will help to determine the difference among the profiles we are comparing. These parameters are: $(i)$ the cross-post similarity value obtained from the methodology described in Section \ref{sec:Cross_posts_methodology}, $(ii)$ the type of content associated to the cross-posts according to the category assigned by the FB API to the posts, $(iii)$ the website associated to the urls contained within TW version of the cross-posts (i.e., tweets).

Due to lack of space we will perform this analysis for the cross-posts shared between FB and TW that, as Table \ref{tab:perc_of_preferences} depicts, represent 66\% of the total cross-posts, which increases to more than 75\% if we also consider the cross-posts that appear in the 3 OSNs (thus also in FB and TW).

\subsection{Cross-Posting behaviour based on Preference}

We create three groups of users according to the results obtained in the previous section (see Table \ref{tab:user_preference_}). The first group, referred to as \textit{TW-favourite}, is formed by the 102 users that show a strong preference for TW. The second group, referred to as \textit{FB-favourite} is formed by the 75 users showing a clear preference for FB. Finally, the last group is formed by the 95 ''agnostic" users that do not show any strong preference, and we refer to it as \textit{Agnostic}. Next we characterize the cross-posting pattern for these four groups based on the three characteristics introduced at the beginning of this section.

Figure \ref{similarity_preference_FBTW} shows the CDF for the cross-post similarity across the users in the three groups. The results show that the users with a strong preference for TW publish more similar cross-posts between FB and TW than the users that prefer FB. In median \textit{TW-favourite} group achieves an average similarity close to 80\%, while \textit{FB-favourite} and \textit{Agnostic} groups reach a median similarity a bit higher than 70\%. 

In order to classify the type of content embedded in the posts we rely on the content type assigned by the Facebook API to each post that can be: text, link, video or photo. It must be noted that posts classified as photo or video by FB API may not include the video or photo in TW, but a link to them. Figure \ref{post_type_preference_FBTW} shows a bar plot presenting the average portion of each type of content published within each group. We observe a substantial difference between \textit{FB-favourite} and \textit{TW-favourite} groups. Users that prefer FB attach photos to half of the cross-posts. Even more, users in this group are the ones that publish a larger portion of videos. In contrast, \textit{TW-favourite} group includes users that publish much less photos and videos  (36\% and 7\% in average, respectively), but much more posts including only text (35\% in average). The agnostic users ranges in between \textit{FB-favourite} and \textit{TW-favourite}.

Finally, we want to find what are the sites more frequently linked from the cross-posts. For this we rely on the urls included in the TW version of the cross-posts (i.e., tweets)\footnote{TW usually employs short urls. Hence, to obtain the website behind the short urls we had to reverse the process and obtain the original urls from the short urls using ``Expand url portal" (http://expandurl.appspot.com/expand?url=http)}

We have found that the most popular websites linked from cross-posts are actually OSNs. In particular, the most linked sites are Facebook, Twitter, Youtube and Instagram. It must be noted that a link to those websites refers in most of the cases to some content (e.g. photo, video, etc) stored in that OSN. Based on these results we analyze the portion of urls linking to those four sites and we group together the remaining urls in a category referred to as \textit{Other}.

Figure \ref{short_url_preference_FBTW} shows a bar plot depicting the average portion of posts including a url linking to Facebook, Twitter, Youtube, Instagram, and Other. Again the results show different patterns for users preferring FB and users preferring TW. For users in the former group 60\% of their urls point to one of the four OSNs, with a clear preference for FB (26\%) and TW (22\%) and a negligible presence of Instagram. In contrast, for users in \textit{TW-favourite} group more than 60\% of the urls link to websites different than the four main OSNs. However, among the OSNs, Instagram is the most popular one (22\% of the urls) while the number of urls for FB is negligible.  Agnostic users are the users including a larger portion of urls to ``Other" websites (almost 70\%).

\emph{In a nutshell, the cross-posting profile of a \textit{TW-favourite} user is as follows: $(i)$ she has a higher similarity for the cross-posts, $(ii)$ she publishes more textual content than audiovisual content, and $(iii)$ she links more frequently websites different than OSNs, but across OSNs it mostly link content in Instagram. In contrast, the profile of a \textit{FB-favourite} user is as follows: $(i)$ she mostly publishes audio-visual content, and $(ii)$ she mostly contains ursl linking content stored on major OSNs, especially stored in FB and TW. Finally agnostic users show an intermediate behaviour between the \textit{TW-favourite} profile and the \textit{FB-favourite} profile.}

\begin{figure*}[t]
	\centering
	\subfigure[{CDF of cross-post similarity}]{\includegraphics[width=0.32\textwidth, height=0.25\textwidth]{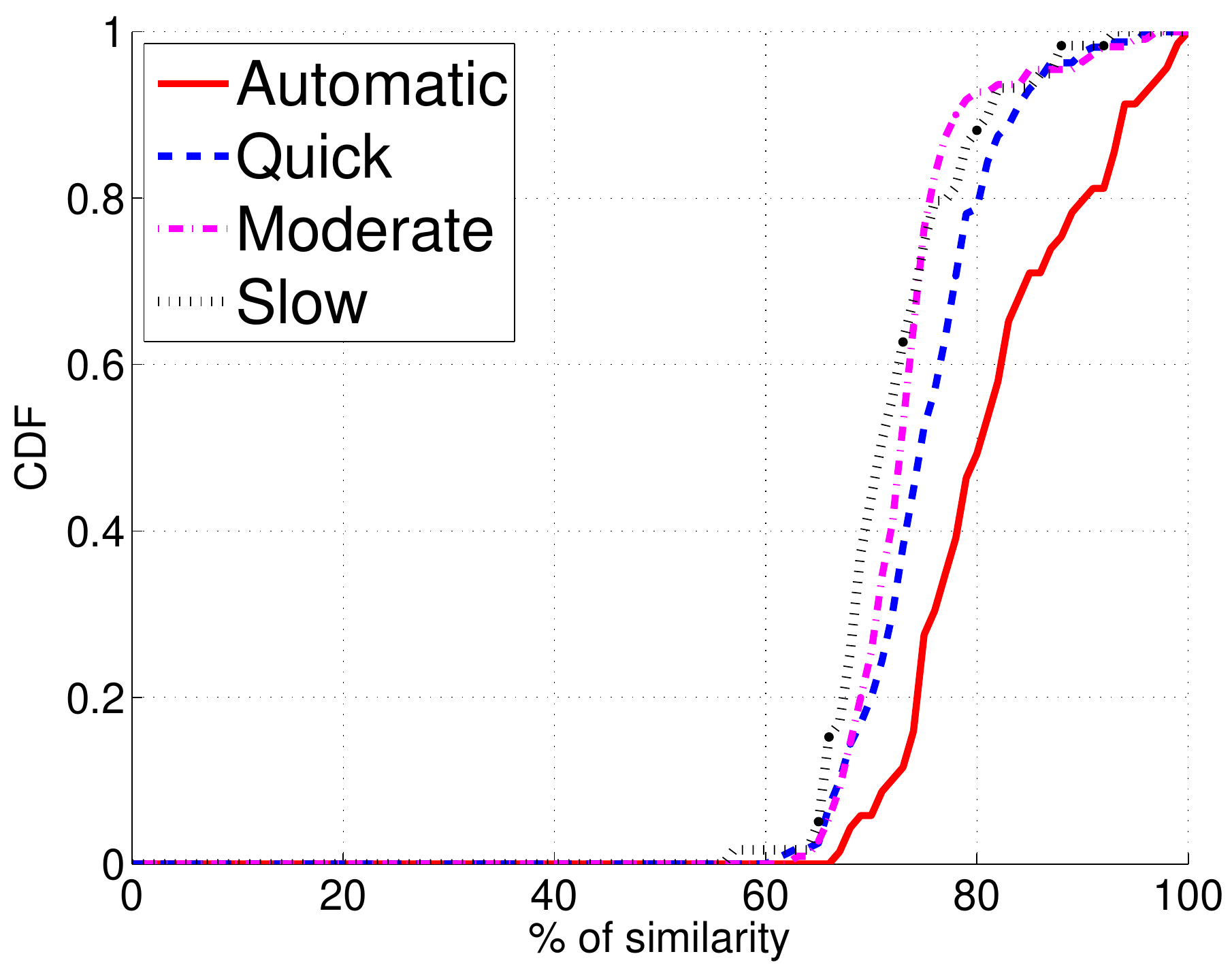}
	\label{similarity_time_FBTW}}
	\subfigure[{Type of post}]{\includegraphics[width=0.32\textwidth, height=0.26\textwidth]{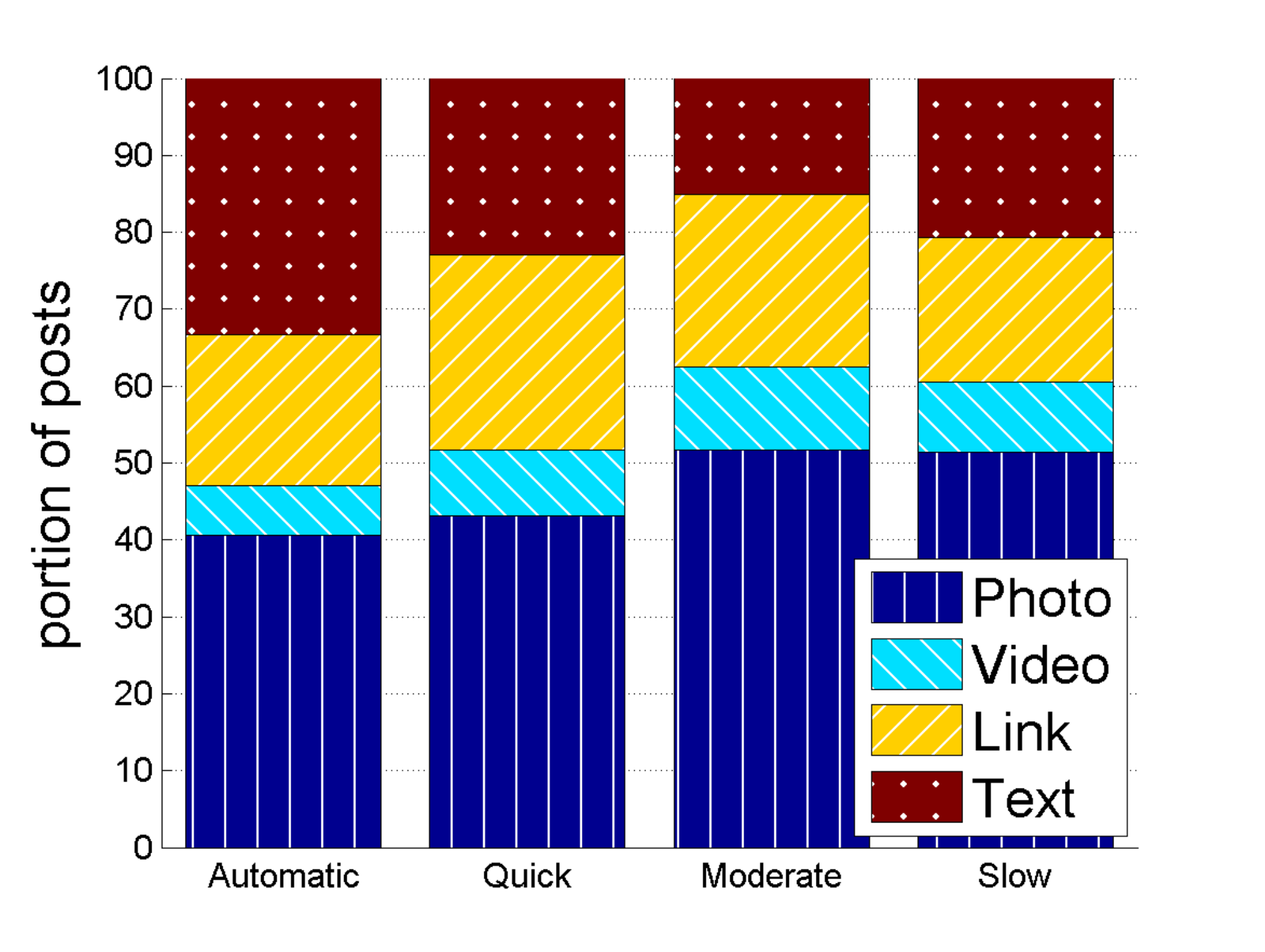}
	\label{post_type_time_FBTW}}
	\subfigure[{Website associated to urls}]{\includegraphics[width=0.32\textwidth, height=0.26\textwidth]{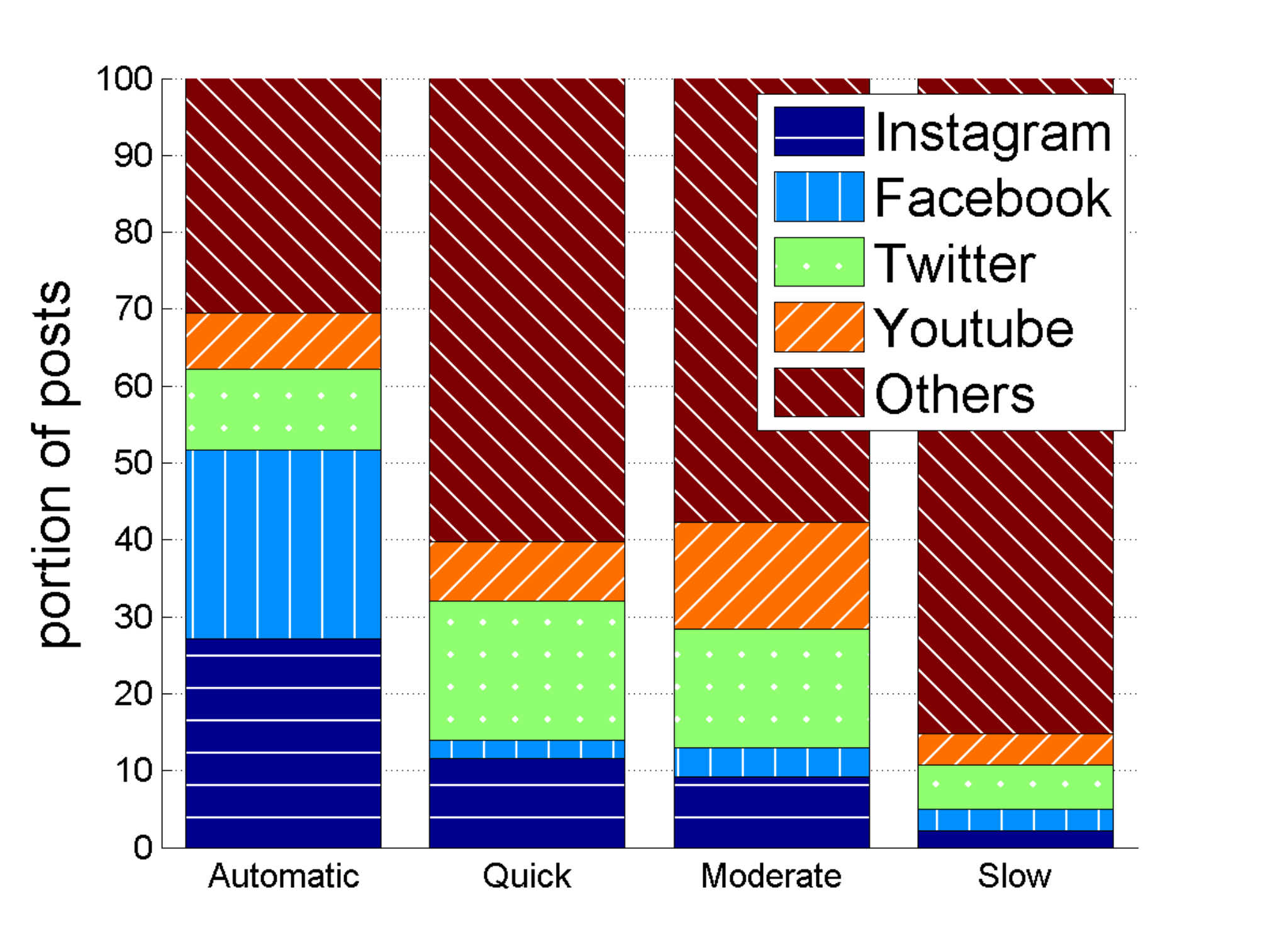}
	\label{short_url_time_FBTW}}
	\vspace{-0.3cm}
	\caption{Cross-posting behaviour characterization based on inter-posting interval.}
\vspace{-0.4cm}
\end{figure*}

\subsection{Cross-Posting behaviour based on Inter-Posting Interval}
We have shown that professional users present different cross-posting pattern depending on the preferred OSN. Similarly, in this section we explore whether the inter-posting interval time reveals different cross-posting profiles as well. Towards this end we group the professional users in our dataset based on the inter-posting interval between FB and TW (independently of the direction). In order to create the groups we apply the K-means clustering algorithm \cite{k_mean} using as parameter the median inter-posting interval of each user. We use this mechanism to discover all the groups, except one that we form manually.

The reason for creating a manual group responds to the fact that more than 5\% of the posts in our data set are published at the same time in two OSNs, and this portion grows to 7\% if we only consider the cross-posts between FB and TW. Even more, 30\% of the cross-posts between FB and TW are posted in both OSNs in a timer interval lower than 10 seconds. Therefore, we thought that there might be a relevant number of users that publish most of their \textsl{FB-TW} cross-posts (in either direction) in less than 10 seconds. We make the assumption that for a human-being is unlikely to manually publish a post both in FB and TW in so short gap. Therefore, we consider that if a post is published in both OSNs in a time interval lower than 10 seconds, it means that the user (e.g., the community manager managing the OSNs accounts of the user) is utilizing some automatic tool to upload the cross-post. Actually, there is quite a few tools that provide this feature [Argyle Social\footnote{\url{http://www.argylesocial.com/}}, dlvr.it\footnote{\url{https://www.dlvr.it/}}, bufferapp\footnote{\url{https://www.bufferapp.com/}}].
Based on this discussion, we have manually formed a group that includes those users that have published more than 1/2 of their cross-posts in less than 10 seconds both in FB and TW. The group is formed by 69 professional users from our dataset. We refer to this group as $Automatic$ since the users forming it are making an intensive use of automatic tools to perform its cross-posting activity.

\begin{table}[t]
  \centering
  \scriptsize
  \vspace{-0.2cm}
  \caption{Average and maximum cross-posting interval within $Automatic$, $Quick$, $Moderate$ and $Slow$ groups.}
  \vspace{-0.3cm}
    \begin{tabular}{l|ccc}
    \hline
      Groups   & \#Users & Avg(inter-posting interval) & Max. \\
    \hline
    Automatic & 69   &   2.3 Minutes  &  68 Minutes\\ 
    Quick & 159  &   8.3 Minutes  &  87 Minutes\\
    Moderate & 109  &    2.35 Hours  & 5.1 Hours \\
    Slow & 59   &    13.5 Hours  & 6.4 days \\
    \hline
    \end{tabular}%
  \label{tab:time_groups_info}%
	\vspace{-0.45cm}
\end{table}%

After creating this group, we run the K-means algorithm \cite{k_mean} to classify the remaining users according to their median inter-posting interval time. We have found 3 clusters whose characteristics (number of users, average and maximum inter-posting interval) along with the characteristics of the $Automatic$ group are depicted in Table \ref{tab:time_groups_info}. We refer to these groups as $Quick$, $Moderate$ and $Slow$.  In a nutshell: $(i)$ users in the $Quick$ group publish their cross-posts in both OSNs in the order of minutes,  $(ii)$ users in the $Moderate$ group take more than 2 hours, and $(iii)$ users in the $Slow$ group takes a gap of more than 13 hours between the publication in the first OSN and the time they republish the post in the second OSN.

Following, we analyze the behaviour of the users in terms of similarity, type of content, and links to websites.

Figure \ref{similarity_time_FBTW} shows the CDF for the cross-post similarity across the users in the different groups for \textsl{FB-TW}. We find a very interesting pattern that correlates the similarity to the inter-posting interval. Basically, the shorter the inter-posting interval the higher the similarity across the cross-posts. This actually is reasonable since if you publish the same information in two OSNs within a very short of time (e.g. $<10s$) the posts are going to look very similar. However, if you post something in TW (FB) and you republish the same information in FB (TW) after some few hours it is more likely that you introduce some change. Finally, we do not observe any relevant difference between the $Moderate$ and the $Slow$ since the capacity of modifying the post is the same for the associated intervals to these groups.

Figure \ref{post_type_time_FBTW}  show the portion of posts belonging to each content category according to the type assigned by FB API. Again we observe an interesting correlation between inter-posting time and type of published content. As the inter-posting interval increases the portion of cross-posts associated to audiovisual content (i.e., photos and videos) also increases, while the combination of more textual content (i.e., pure text +  links) decreases. In particular, the quickest category, i.e., $Automatic$ group, publishes around 45\% of audiovisual posts, which increases to 50\% for $Quick$ group and to 60\% for $Moderate$ and $Slow$ groups. Therefore, it seems that automatic tools are employed less frequently to upload videos and photos.


Figure \ref{short_url_time_FBTW} shows a bar plot depicting the average portion of posts including a cross-post linking to Facebook, Twitter, Youtube, Instagram, and Other. We observe a large divergence for the results across the different groups. First, almost 70\% of the urls included in cross-posts of users belonging to $Automatic$ group are linking to one of the four main OSNs, with a strong presence for FB (25\%) and Instagram (27\%). All the remaining groups contains more urls linking ``Other" websites than urls linking the four OSNs. 
It is particularly interesting the very marked pattern of the users in the $Slow$ group for which only 15\% of the links go to OSNs.

\emph{In a nutshell, as the inter-posting interval decreases: $(i)$ the similarity of cross-posts increases, $(ii)$ the portion of audiviousal content attached to cross-posts decreases, $(iii)$ and a larger portion of urls included in cross-posts refers to major OSNs sites.}


\vspace{-0.2cm}
\section{Related Work}
\label{sec:rel_work}

There exist several works that have studied the graph and connectivity properties of Facebook, \cite{Facebook1,Facebook2,gjoka2010walking}, Twitter \cite{Twitter_WWW2010,cha2010measuring}, and Google+ \cite{g+_imc_cha,schioberg2012tracing}. In our previous study \cite{H2H_report}, we perform a head-to-head comparison of FB, TW and G+ in terms of connectivity and activity across different group of users.
In addition, there are other works in the literature that compare two or more OSNs based on their graph properties \cite{Reza_IEEENetwork,Mislove_IMC07} or other social aspects \cite{www_mostafa_Bootstrapping}. However, these works do not consider the same users in the different OSNs for their analysis since their goal is to characterize OSNs at a macroscopic level.

There are only few works that try to characterize the behaviour of the same user or group of users across different OSNs. The main reason is that it is not an easy task to identify and collect the information of the same users across different system and, in addition, it requires to have one data collection tool for each system. There are some few tools and platforms available in the market \cite{Socialbakers,pagedatapro} that provide some few information (for free) of a given user across different OSN. However, that information is usually limited to the number of followers, the number of published posts, aggregated engagement and/or popularity trends. Therefore, these tools do not provide enough detail on the activity of a user to perform a comprehensive analysis of its behaviour in different OSNs.

Nevertheless, some few studies in the literature have analyzed the behaviour of the same users across different OSNs \cite{mytweet, A_tale_two_sites}.
Authors in \cite{Archival} compare 195 users from the archival community and study their activity pattern in TW and FB.  This is a small-scale study based on 2,926 links to external documents. In \cite{A_tale_two_sites}, we find again a comparative analysis for users having accounts in FB and TW. This work studies the behaviour of 300 users from a psychological perspective and the results reveal a correlation between end-users personality and their use of FB and TW. Finally, the most similar work to our paper is a very recent study \cite{Of_Pins_Tweets} that compares the behaviour of 30,000 regular users across TW and Pinterest. Although this study similar in spirit to our work, we differ from \cite{Of_Pins_Tweets} since we are focusing in professional OSN players instead of regular users, and we are comparing TW, FB and G+ instead of TW and Pinterest.

\section{Conclusions}
\label{sec:conclusions}
\vspace{-0.3cm}
This paper presents the first large-scale measurement-based characterization of the cross-posting activity for OSN professional users across FB, TW and G+. We have used a simple yet efficient methodology that is able to determine with an accuracy of 99\% whether two posts, even from different OSNs, contains the same information, and if so classify them as cross-post. We have used that methodology to classify more than 2M posts published for 616 professional publishers with active accounts in FB, TW and G+. Following we list the main outcomes of the paper. First, we have demonstrated that professional users frequently publish the same information in at least two OSNs, especially in the case of FB and G+. Although professional users in TW present a low portion of cross-posts, the fact that they are very active implies that in absolute terms we can find quite a lot cross-posts in their TW accounts. Second, a professional user publishes (in median) 70\% of her cross-posts exclusively in FB and TW, and around 15\% in FB and G+. Furthermore, we demonstrated that the cross-posting activity between TW and G+ is negligible. Third, professional users benefit of cross-posting in their TW and FB accounts since they attract 2$\times$ and 30\% more engagement with cross-posts than non-cross-posts, respectively. However, cross-posts in G+ leads to halve the engagement as compared to non-cross-posts. Fourth, professional users equally prefer FB and TW as initial source of information, but they rarely choose G+. Fifth, users with a strong preference for TW present cross-post with a higher similarity (across different OSNs), publish more textual content than photos and videos, and use to include links to websites different than major OSNs. In contrast, users preferring FB publish mainly audiovisual content and a major portion of urls in their cross-posts refer to OSN content. Finally, as the user inter-posting interval time decreases: (i) the similarity of her cross-posts increases, (ii) the portion of audiovisual content attached to her cross-posts decreases as wall, (iii) and a larger portion of urls included in her cross-posts refers to major OSNs sites.
As future direction of this research, we aim to conduct a deeper user-level analysis to understand how different categories of users are vary to each other based on different strategies and metrics such as their level of cross-posting activity.

\section{Acknowledgments}
\vspace{-0.3cm}
\small
This work is partially supported by the European Celtic-Plus project CONVINcE and ITEA3 CAP.
as well as the Ministerio de Economia y Competitividad of SPAIN through the project BigDatAAM (FIS2013-47532-C3-3-P) and Horizon 2020 Programme (H2020-DS-2014-1) under Grant Agreement number 653449.
We would like thank Reza Motamedi, Reza Rejaie, Roberto González and Ruben Cuevas for providing Twitter and Google+ dataset to be used in this study.


\bibliographystyle{IEEEtran}




\end{document}